\documentclass[12pt]{article}

\usepackage[top=1in, bottom=1in, left=1in, right=1in]{geometry}
\usepackage{amssymb}
\usepackage{amsmath}
\usepackage{graphicx}
\usepackage{placeins}
\usepackage{subcaption}
\usepackage{xcolor}
\usepackage{tabls}
\usepackage{appendix}
\usepackage{ragged2e}
\usepackage{enumitem}
\setlist[itemize]{leftmargin=*}
\usepackage{caption}
\usepackage{multirow}

\usepackage{authblk}

\newcommand{\vspaceBefFigCaption}{\vspace*{-0.5cm}}
\newcommand{\vspaceBefTableCaption}{\vspace*{-0.2cm}}
\newcommand{\citep}[1]{\cite{#1}}

\begin{document}

\title{Direct Low-Dose CT Image Reconstruction on GPU 
       using Out-Of-Core: Precision and Quality Study}

\author[1]{M.~Chillar\'{o}n}

\affil[1]{
  {Universitat Politècnica de València},
  {Departamento de Sistemas Informáticos y Computación, Camí de Vera}, 
  {Valencia},
  {46022}, 
  {Valencia},
  {Spain.}
}

\author[2]{G.~Quintana-Ort\'{i}}

\affil[2]{
  {Universitat Jaume I de Castellón},
  {Departamento de Ingeniería y Ciencia de Computadores, Av.~Sos Baynat}, 
  {Castellón},
  {12071}, 
  {Castellón},
  {Spain.}
}
            
\author[1]{V. Vidal}

\author[3]{G.~Verd\'{u}}

\affil[3]{
  {Universitat Politècnica de València},
  {Departamento de Ingeniería Química y Nuclear, Camí de Vera}, 
  {Valencia},
  {46022}, 
  {Valencia},
  {Spain.}
}

\setcounter{Maxaffil}{0}
\renewcommand\Affilfont{\itshape\small}

\maketitle

\begin{abstract}
Algebraic methods applied to the reconstruction of Sparse-view Computed Tomography (CT)
can provide both a high image quality
and a decrease in the dose received by patients,
although with an increased reconstruction time 
since their computational costs are higher.
In our work, 
we present a new algebraic implementation
that obtains an exact solution 
to the system of linear equations that models the problem 
and based on single-precision floating-point arithmetic.
By applying Out-Of-Core (OOC) techniques,
the dimensions of the system can be increased 
regardless of the main memory size
and as long as there is enough secondary storage (disk).
These techniques have allowed to process images of $768 \times 768$ pixels.
A comparative study of our method on a GPU 
using both single-precision and double-precision arithmetic
has been carried out.
The goal is to assess the single-precision arithmetic implementation
both in terms of time improvement and quality of the reconstructed images
to determine if it is sufficient to consider it a viable option. 
Results using single-precision arithmetic approximately halves 
the reconstruction time of the double-precision implementation,
whereas the obtained images retain all internal structures 
despite having higher noise levels.
\end{abstract}



\section{Introduction}
\label{sec:intro}

In the field of computed tomography (CT) reconstruction, 
the time required to obtain the images is a key factor
since in clinical practice images need to be examined in real time 
in order to determine the validity of a study. 
Hence, it is very important to develop time-efficient methods 
with the necessary quality.

Traditional methods based on the filtered back projection (FBP) technique continue 
to be the fastest ones~\citep{fbp}. 
However, they also require the highest radiation dose,
which is an important drawback.
Therefore, the methods employed in practice 
have evolved towards iterative reconstruction (IR)
methods~\citep{willemink2013iterative}. 
Iterative analytical methods seek to alleviate 
the quality-decrease effects of dose reduction 
(whether due to intensity or voltage reduction), 
obtaining images of similar quality to FBP with a lower radiation dose. 
They are based on iterative cycles in which the image is corrected 
by eliminating noise through non-linear processing based on statistical information 
until the optimum quality is obtained with a minimum noise level. 
These corrections can be made either based on the sinogram, 
or based on the image alone, or both.

Most iterative methods have been developed 
by the scanner manufacturers for their own devices.
For example, Philips has developed the iDose4 method, 
Siemens has developed IRIS and SAFIRE, 
and Toshiba Medical Solutions has developed AIDR 3D.
Although 
these methods achieve a significant dose reduction, 
they also increase the time required to obtain the images. 
Nevertheless, 
they have been optimized so that they can compete 
with the FBP method in terms of time constraints. 
For instance, 
a study~\citep{korn2012iterative} showed that 
the IRIS method performs a complete reconstruction 
in an average of 68 seconds versus 25 seconds for the FBP. 
Another study~\citep{moscariello2011coronary} showed that
the number of slices per second is 20 with SAFIRE versus 40 with FPB.
In another one~\citep{funama2011combination},
iDose4 achieved 16 slices per second versus 22 of FPB.
In short, all these methods 
are competitive enough to be used in real time, 
obtaining reconstructions in-situ while the study is being carried out.

On the other side,
another strategy to reduce the radiation dose 
is to use a sparse sampling scheme to acquire the patient's data. 
This would require designing new CT scanners,  
but some prototypes~\citep{chen2018first} prove its feasibility.
This new approach requires algebraic methods 
to obtain the reconstructions, 
since analytical methods do not perform well due to undersampling. 
Algebraic methods can also provide a high image quality, 
allowing a decrease in the amount of dose received by the patient.
However, their computational costs are much higher than those of analytical methods, 
and therefore the reconstruction time is larger.
A previous work~\citep{chillaron2020computed}
showed how a speed of about 1.3 slices per second can be achieved 
when using the dense QR method implemented with Out-Of-Core (OOC) techniques 
on an affordable computer (with no GPU) with double-precision data.
In another work~\citep{quintana2022high}, 
the code was ported for GPU execution and was also deeply optimized, 
achieving approximately 10 slices per second with double-precision data.
This speed is close to those obtained with iterative methods.

On the other hand, 
due to the high quality of the reconstructions,
working with double-precision floating-point arithmetic might not be necessary.
To explore this, we have developed a 
new implementation of the QR method 
using OOC computing techniques applied to CT reconstruction 
with single-precision floating-point arithmetic, 
presented here.
Recall that a double-precision floating-point number requires 64 bits,
a single-precision floating-point number requires 32 bits,
and single-precision arithmetic operations are usually faster.
Moreover, we have carried out a comparative study of 
an implementation using single-precision arithmetic and 
an implementation using double-precision arithmetic.
The goal is to assess both the quality of the images and 
the time required in the reconstructions,
with the final aim of determining 
if the use of single-precision arithmetic is a valid option.

\section{Materials and Methods} 
\label{sec:materials}

To reconstruct CT images with an algebraic approach, we model the problem as:
\begin{equation}
    AX = B
    \label{system}
\end{equation}
where $A = (a_{i,j}) \in R^{M \times N}$ is the weights matrix 
that models the physical scanner,
$a_{i,j}$ is the contribution of the $i$-th ray at the $j$-th pixel,
$M$ is the number of traced rays, and
$N$ is the number of pixels of the image to be reconstructed. 
The sinograms matrix (or vector if only one sinogram is used) is $B$, 
and $X$ is the solution image.
The solution $X$ can be computed according to the following formulae
from matrix linear algebra:

\begin{eqnarray}
    A & = & QR \label{qr} \\
    X & = & R^{-1}(Q^{T}B) \label{solve}
\end{eqnarray}

To solve the problem in Eq.~\ref{system} with a direct algebraic approach, 
we first calculate the QR factorization of $A$ in Eq.~\ref{qr},
where $Q$ is orthogonal (or has orthonormal columns when $A$ is not square), 
and $R$ is upper triangular. 
Then, to reconstruct the images, Eq.~\ref{solve} is used.
The operation in Eq.~\ref{solve} can be broken down into two computations:
$C=Q^TB$ and $X=R^{-1}C$.
Note that matrix $Q$ is not explicitly built
since it is more expensive and requires more storage.
Analogously, since $R$ is upper triangular,
the inverse $R^{-1}$ is not explicitly built.
It is important to observe that Eq.~\ref{qr} must be recomputed only if the
coefficient matrix $A$ is different.
That is, if two linear systems of equations share the coefficient matrix $A$,
the computation in Eq.~\ref{qr} can be reused in the second system,
thus saving a large amount of computation.

\subsection{Resolution on data stored in main memory}

When all the data
(matrix $A$, matrix $B$, temporal data, etc.)
are dense and fit in the main memory of the computer,
the above process (Eq.~\ref{qr} and~\ref{solve}) 
can be easily computed 
with just three calls to the corresponding routines of 
the LAPACK and BLAS libraries:
A first call to \texttt{dgeqrf} for computing the QR factorization,
a second call to \texttt{dormqr} for applying $Q^T$ (computation of $C = Q^TB$),
and a final call to \texttt{dtrsm} for computing $R X = C$.
To achieve high performances,
optimized versions of the LAPACK and BLAS libraries,
such as Intel MKL or NVIDIA cuSOLVER,
must be employed.

\subsection{Resolution on data stored in secondary storage}

However, when the linear system is large enough,
the data (matrix $A$, matrix $B$, etc.)
will not fit in main memory.
For instance, 
to generate an image of $512 \times 512$ pixels,
the coefficient matrix $A$ requires about 544 GB 
when working in double precision.
To generate an image of $768 \times 768$ pixels,
the coefficient matrix $A$ requires about 2.6 TB
when working in double precision,
which is an impressive as well as expensive amount of main memory,
considering the current main memory sizes of 
personal computers (about 16 or 32 GB) and servers (about 128 or 256 GB).

Therefore,
when the linear system is so large that it does not fit in the main memory
and must be instead stored in secondary storage (disk),
Out-Of-Core techniques can be employed to be able to process it.
By using these techniques with a careful programming,
performances nearly similar 
to those obtained when processing data in main memory
can be obtained with the only drawback of a much more complex code.
We have developed and assessed an Out-Of-Core implementation
that uses two techniques to achieve high performance:
The first technique is the use of a part of the main memory 
as a cache of the data in the secondary storage (disk),
thus effectively reducing the amount of data being transferred.
The second technique is to overlap the computations 
in the computing device (GPU) with the disk I/O access,
thus completely hiding the data access.
Our software uses the high-performance NVIDIA cuSOLVER library
to carry out the computational tasks.
Although the final software is much more complex,
performances can be very high
without having a very large and expensive main memory.

To achieve this,
the initial matrices are partitioned into square data blocks
(except for the left-most and bottom-most blocks
if the matrix dimensions are not multiple of the block dimensions).
Then, the above process (Eq.~\ref{qr} and~\ref{solve}) 
is broken down into many basic tasks that are applied to those data blocks.
Every basic task is of a given computational type (operation) and
can have several input, output, and input/output operands.
To simplify the programming,
a runtime executes all these tasks
taking into account the data cache and 
overlapping the computation of one task with 
the transfer of operands of the following tasks.

For instance, let us suppose that 
the operation $R X = C$ must be computed.
Let us partition matrices $R$, $X$, and $C$ in the following way:
\[
  R = \left( 
      \begin{array}{ccc}
        R_{00} & R_{01} & R_{02} \\ 
               & R_{11} & R_{12} \\ 
               &        & R_{22} \\ 
      \end{array}
      \right),
  \qquad
  X = \left( 
      \begin{array}{c}
        X_{0} \\ 
        X_{1} \\ 
        X_{2} \\ 
      \end{array}
      \right),
  \qquad
  C = \left( 
      \begin{array}{c}
        C_{0} \\ 
        C_{1} \\ 
        C_{2} \\ 
      \end{array}
      \right)
\]
To perform the operation $R X = C$,
the following tasks must be computed:

\begin{enumerate}
\item  $R_{22} X_2 = C_2$
\item  $C_1 = C_1 - R_{12} X_2$
\item  $R_{11} X_1 = C_1$
\item  \ldots
\end{enumerate}

As can be seen above, 
the initial operation $RX=C$
has been decomposed into several tasks of two types.
For instance, the first and third tasks compute 
the solution of upper-triangular systems,
whereas the second task computes a matrix-matrix product.

The runtime executes all the above tasks one by one.
Before executing the first task,
it must load all its operands into the block cache in main memory from disk.
While executing the first task,
the runtime brings in advance the operands for the second task (and next ones) 
from disk into main memory if they are not already in the block cache.
While executing the second task,
it brings in advance the operands for the third task (and next ones) 
if they are not in main memory.
In this way, computations and disk I/O are overlapped.
Of course,
when working with real data, the total number of blocks 
are several hundreds or even thousands,
and therefore the number of tasks is obviously much larger.


A more detailed explanation on the resolution method and its implementation for both CPU and GPU can be found in previous works~\citep{chillaron2020computed,quintana2022high}. 
These perform an in-depth analysis of 
both the time efficiency obtained using double precision and 
the quality of the reconstructed images.
Since the QR method using double precision obtains very low residuals 
and the image quality metrics show very small errors, 
precision might be reduced
without a drastic drop in the quality of the reconstructed images.

In this document 
we introduce and assess a new implementation
of the algebraic method for reconstructing CT images
that employs single-precision floating-point numbers (32 bits)
instead of double-precision floating-point numbers (64 bits).
This change can allow to perform the computations faster,
as well as reducing the size of the dataset.
Moreover, our implementation works
on linear systems of any size stored in the disk
and regardless of the memory size.

\section{Experimental Results} 
\label{sec:results}

The experiments for this work have been carried out 
on a server called Alinna. 
It contains two AMD EPYC 7282 processors (32 cores),
a main memory of 768 GiB, and 
an NVIDIA A100 GPU with 40 GiB of RAM. 
Our software uses the GPU as the computing device, 
and the libflame~\citep{CiSE09}, cuSOLVER, and cuBLAS libraries.

\subsection{Dataset COVID-CT-MD}
\label{sec:covid}

The CT images used in this part of our experimental study have been selected 
from the COVID-CT-MD data set~\citep{afshar2021covid}, 
which includes a collection of real thorax CT images obtained 
from patients without pathologies, 
patients with COVID-19-related pneumonia, and 
patients with community-acquired pneumonia (CAP).
In our experiments,
the maximum number of simultaneous slices has been set to 2048, 
since it is enough for a full-body scan. 
In a real clinical study, 
the number of slices depends on the size of the area to be analyzed and 
the acquisition settings.

\subsubsection{Simulated scanner parameters}

For this study, 
the resolution of the reconstructed images is the standard one: $512 \times 512$ pixels. 
The scanner has $1025$ detectors, and $260$ views or projections are used, which is the minimum number that guarantees that matrix A has full rank, as shown in previous studies~\citep{chillaron2018ct}. 
When using the QR factorization, having a full-rank matrix is mandatory.
The minimum number of projections is determined 
by the dimensions of the problem. 
In this case, the number of rows in the matrix $A$ should be 
at least equal to the number of columns. 
Since it has $262,144$ columns (the image resolution in pixels), 
the number of rows can be at least equal to that number. 
Given that the number of detectors is 1025, 
the minimum number of projections is 256. 
However, we use 260 to ensure that the rank is complete. 
If the problem was still ill-conditioned, this number could be increased.

With this configuration, 
the sinograms or projection data have been simulated using the selected images from the dataset.
Table~\ref{tab:scanner} details the parameters of the simulation.
The simulation method used is Joseph's forward projection method~\citep{Joseph}. 

\begin{table}[ht!]
\centering
  \begin{tabular}{|c|c|} \hline
    \textbf{Parameter}                & \textbf{Value} \\  \hline
    Source trajectory                 & $360^\textrm{o}$ circular scan \\  \hline
    Source-to-isocenter distance      & 75 cm \\  \hline
    Source-to-detector distance       & 150 cm \\  \hline
    X-ray source fan angle            & $30^\textrm{o}$ \\  \hline
    Number of detectors               & 1025 \\  \hline
    Pixels of the reconstructed image & $512 \times 512$ \\  \hline
    Number of projections             & 260 \\ \hline
  \end{tabular}
  \vspaceBefTableCaption
  \caption{\small{Simulated fan-beam scanner parameters.}}
  \label{tab:scanner}
\end{table}

\subsubsection{Precision and time efficiency}

Table~\ref{tab:residual_and_times} includes 
both the precision results and the computational times for 
both double-precision implementations 
and single-precision implementations.
To assess the precision and quality of the solution, 
the $|| AX-B ||_{F} / || B ||_{F}$ residual has been computed
after solving the linear system $AX = B$.
As can be seen in the table,
the residual of the single-precision implementation is about $10^{-6}$,
whereas that of the double-precision implementation is about $10^{-12}$,
that is, the single precision is about 6 orders of magnitude higher (worse).
This increase is very significant at a numerical level, 
so it will be necessary to analyze how it affects the reconstructed images,
which will be done afterwards.
Anyway, it is important to note that 
the residuals obtained with iterative algebraic methods 
using double-precision arithmetic
such as the Least Squares QR (LSQR) method are about $10^{-6}$.
Therefore, our new implementation for single-precision data 
would be competitive with previous works with iterative methods 
working with double-precision data~\citep{CHILLARON20171195,E.Parcero}.

Table~\ref{tab:residual_and_times} also contains the time required to reconstruct 
a collection of 2048 slices (computation in Eq.~\ref{solve}),
as well as the total time for computing the QR decomposition
(computation in Eq.~\ref{qr}).
As can be seen,
the resolution time is reduced by 49\% 
(from $2.5$ minutes when using double precision
to $1.3$ minutes when using single precision). 
This time difference might seem not too significant, 
but in a clinical environment 
it means that more patients per day can be treated.
The improvement in the percentage of time 
for computing the QR factorization is smaller (29\%), 
saving about 16 minutes of computation (from 57 to 41 minutes). 
As previously said, this step is not so critical
since it can be reused many times.

\begin{table}[ht!]
  \centering
  \begin{tabular}{|c|c|c|c|} \hline
    \textbf{Precision} & 
    \textbf{Residual} &
    \textbf{Time of resolution} &
    \textbf{Time of QR} \\ \hline
    Double &  $7.52 \cdot 10^{-12}$ &  $148.1$ &  $3\ 439.7$ \\ \hline
    Single &  $2.64 \cdot 10^{-6}$  &   $76.1$ &  $2\ 444.9$ \\ \hline
  \end{tabular}
  \vspaceBefTableCaption
  \caption{Residuals and times in seconds 
           for both single-precision arithmetic and double-precision arithmetic 
           when reconstructing 2048 simultaneous slices.
           The resolution time includes the time in Eq.~\ref{solve},
           whereas the QR time includes the time in Eq.~\ref{qr}.
           }
  \label{tab:residual_and_times}
\end{table}

\begin{figure}[ht!]
  \centering
  \includegraphics[width=0.60\linewidth]{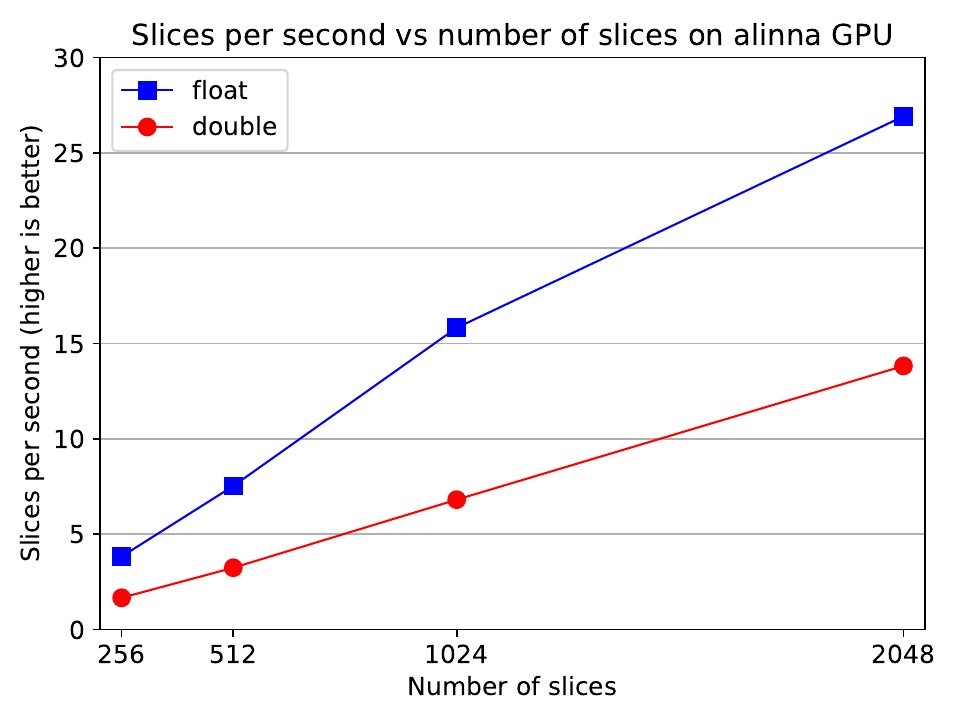} 
  \vspaceBefFigCaption
  \caption{Slices per second versus number of slices being simultaneously computed
  for single-precision implementations (\texttt{float})
  and double-precision implementations (\texttt{double})
  on Alinna GPU.
  }
  \label{fig:slices_per_second}
\end{figure}

Figure~\ref{fig:slices_per_second} shows the speed in slices per second
with respect to the number of slices being simultaneously computed
for both single-precision implementations
and double-precision implementations.
%
%
Note that in both cases the speed clearly increases as the number of slices increases, 
which can be very useful to generate many more slices 
with only a slightly higher overall time.
As can be seen,
the implementation that works on single-precision data
achieves a speed twice as large 
as the implementation that works on double-precision data.
In addition, 
recall that the single-precision implementation 
also requires half the storage.
Therefore, the speeds obtained by our methods are competitive 
with current traditional and analytical methods.

\subsubsection{Image quality}

To objectively analyze the images obtained, 
several quality metrics will be used, 
which measure both the error in the pixel values and 
the conservation of the internal structures 
of the images~\citep{hore2010image}. 
They are the following ones:

\begin{itemize}

\item
Signal-to-Noise Ratio (SNR):
It measures how the original image is affected by noise
and is calculated by Eq.~\ref{eq:snr}. 
Note that in these equations 
$M$ and $N$ are the number of rows and columns of pixels of the image, 
respectively.

\begin{equation}\label{eq:snr}
  \textrm{SNR} = 10 \log_{10}\frac{\sum_{i=0}^{M-1}\sum_{j=0}^{N-1}(I_0(i,j))^2}{\sum_{i=0}^{M-1}\sum_{j=0}^{N-1}(I_0(i,j)-I(i,j))^2}
\end{equation}
\item
Peak-Signal-to-Noise ratio (PSNR): 
It measures the power of the signal relative to noise in high-intensity areas. 
It uses the Mean Square Error (MSE), 
which is calculated according to Eq.~\ref{eq:mse}
and represents the mean of the squared error between the reference image $I_{0}$ 
and the reconstructed image $I$ ($X$ in our equations).
Once the MSE is calculated, 
the PSNR can be computed 
according to Eq.~\ref{eq:psnr}, where $\textrm{MAX}$
is the maximum value that a pixel can take. 

\begin{equation}\label{eq:mse}
  \textrm{MSE} = \frac{1}{MN}\sum_{i=0}^{M-1}\sum_{j=0}^{N-1}(I_0(i,j)-I(i,j))^2
\end{equation}
\begin{equation}\label{eq:psnr}
  \textrm{PSNR} = 10 \log_{10} \frac{\textrm{MAX}_{I_0}^2}{\textrm{MSE}}
\end{equation}

\item
Structural Similarity Index (SSIM): 
It measures the level of preservation of the internal structures and edges 
of the images. 
It is applied through windows of fixed size, and 
the difference between two windows $x$ and $y$ corresponding to the two images to be compared 
is calculated using Eq.~\ref{eq:ssim}.
In this equation, 
$\mu_{x}$ and $\mu_{y} $ denote 
the average value of the windows $x$ and $y$, respectively.
Moreover, 
$\sigma^{2}_{x}$ and $\sigma^{2}_{y}$ denote the variances of both windows,
$\sigma_{xy}$ denotes the co-variance between the windows, 
and $c_1$ and $c_2$ are two stabilizing variables 
dependent on the dynamic range of the image. 

\begin{equation}\label{eq:ssim}
  \textrm{SSIM} = \frac{(2\mu_x\mu_y+c_1)(2\sigma_{x,y}+c_2)}{(\mu_x^2+\mu_y^2+c_1)(\sigma_x^2+\sigma_y^2+c_2)}
\end{equation}

\end{itemize}

In all these metrics, higher is better.
Also note that the maximum value of the SSIM metric is 1.


Table \ref{tab:quality} 
shows the mean values of the previous metrics 
for the entire collection of 2048 images described above.
Note that both SNR and PSNR results are notably 
higher when double-precision data is used.
On the other hand,
it is important to note that 
these results with single precision are comparable to 
iterative algebraic methods such as LSQR working with double precision.
Finally,
mean values of SSIM with single precision and double precision hardly differ. 
Taking into account that the maximum value of SSIM that can be obtained is 1, 
which means that at a visual level
all the image structures are perceived without error, 
it can be said that while visually perfect images are obtained with double precision, 
with single precision the visual error obtained is minimal.

\begin{table}[ht!]
  \centering
  \vspace{0.3cm}
  \begin{tabular}{|c|c|c|c|} \hline
    \textbf{Precision} & \textbf{SNR} & \textbf{PSNR} & \textbf{SSIM} \\ \hline
    Double             & 197.72       & 210.32        & 1             \\ \hline
    Single             &  31.18       &  44.06        & 0.99998       \\ \hline
  \end{tabular}
  \vspaceBefTableCaption
  \caption{Mean image quality of the 2048 slices.}
  \label{tab:quality}
\end{table}

\begin{figure}[ht!]
  \centering
  \begin{tabular}{ccc}
    \includegraphics[width=0.30\linewidth]{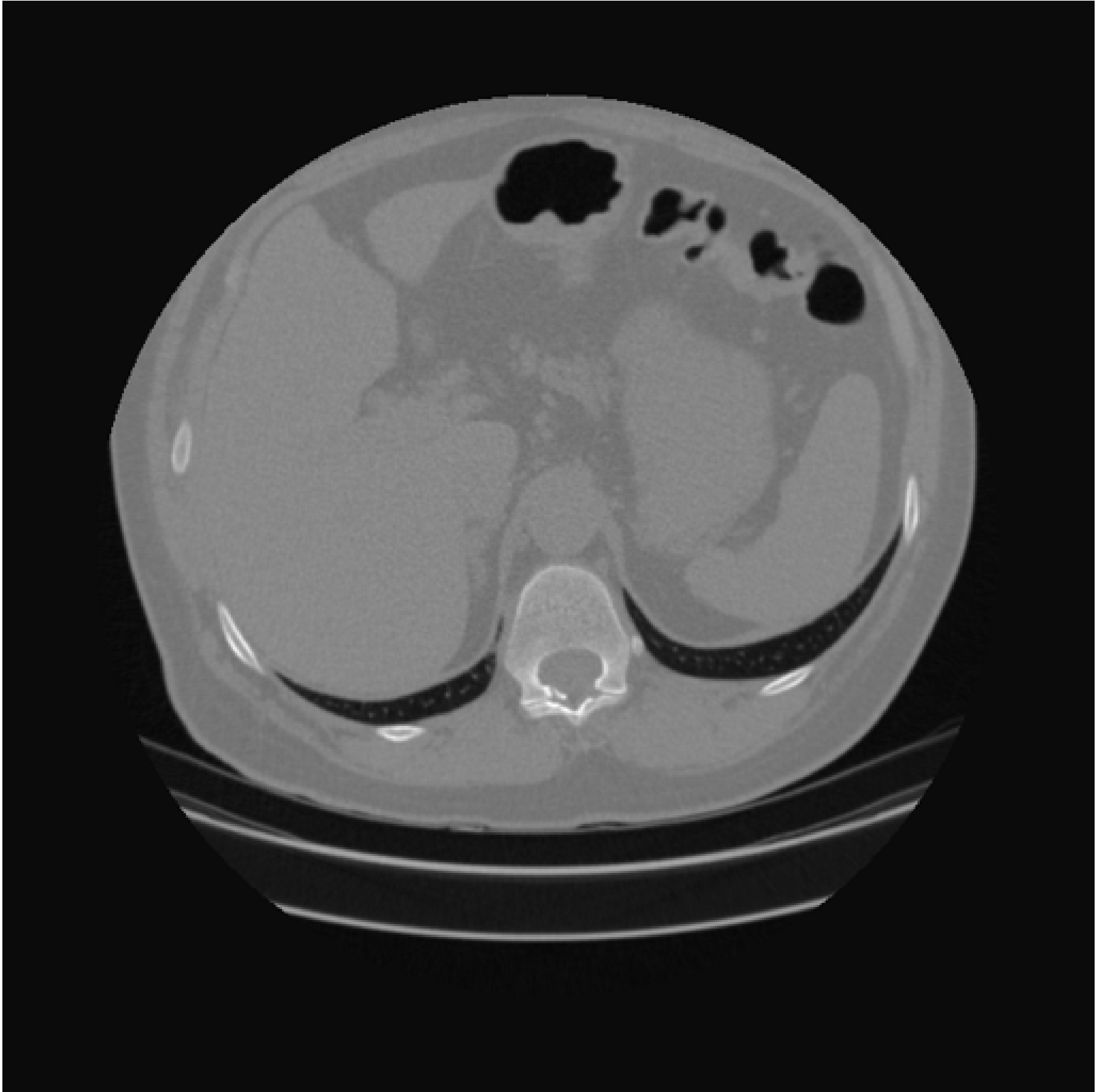} &
    \includegraphics[width=0.30\linewidth]{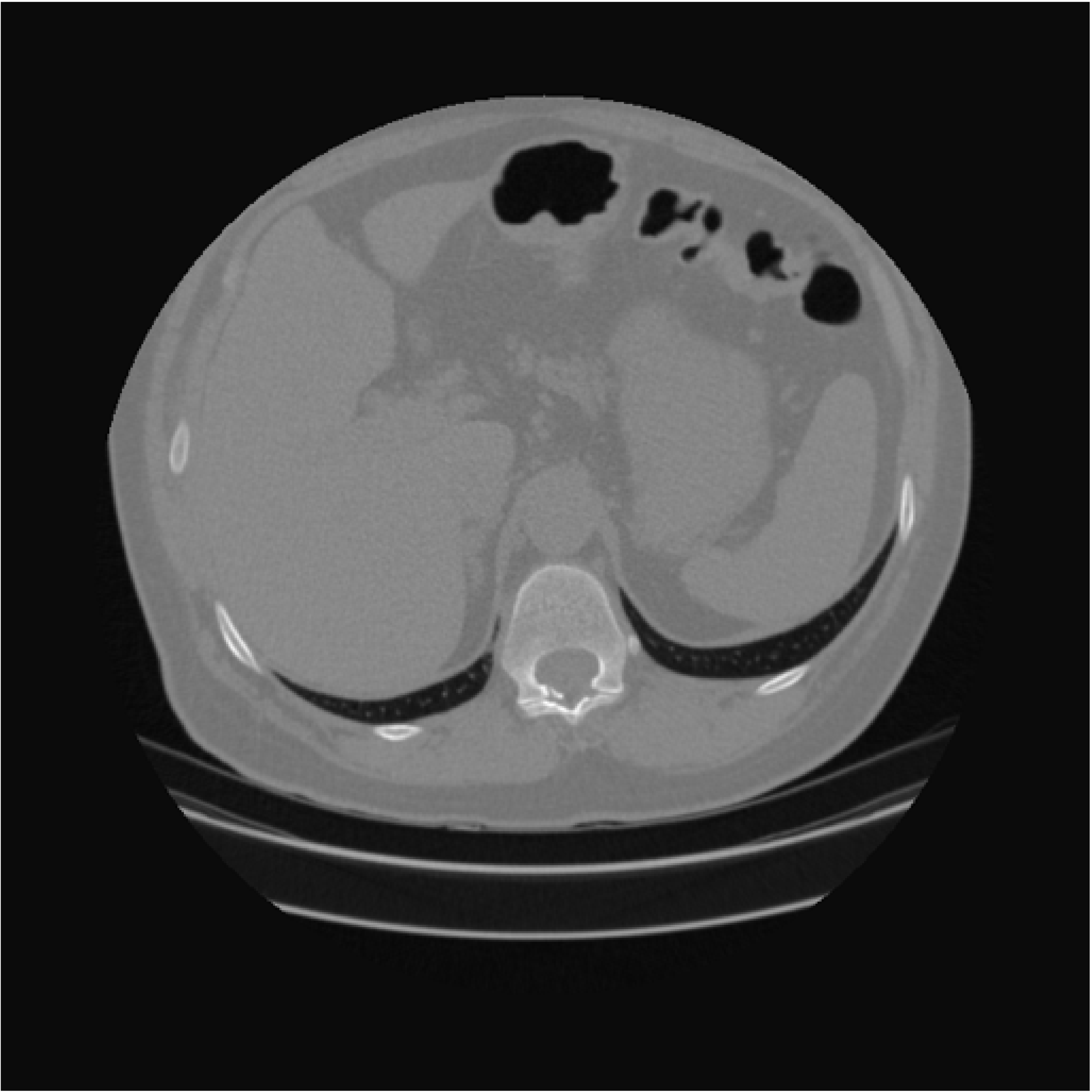} &
    \includegraphics[width=0.30\linewidth]{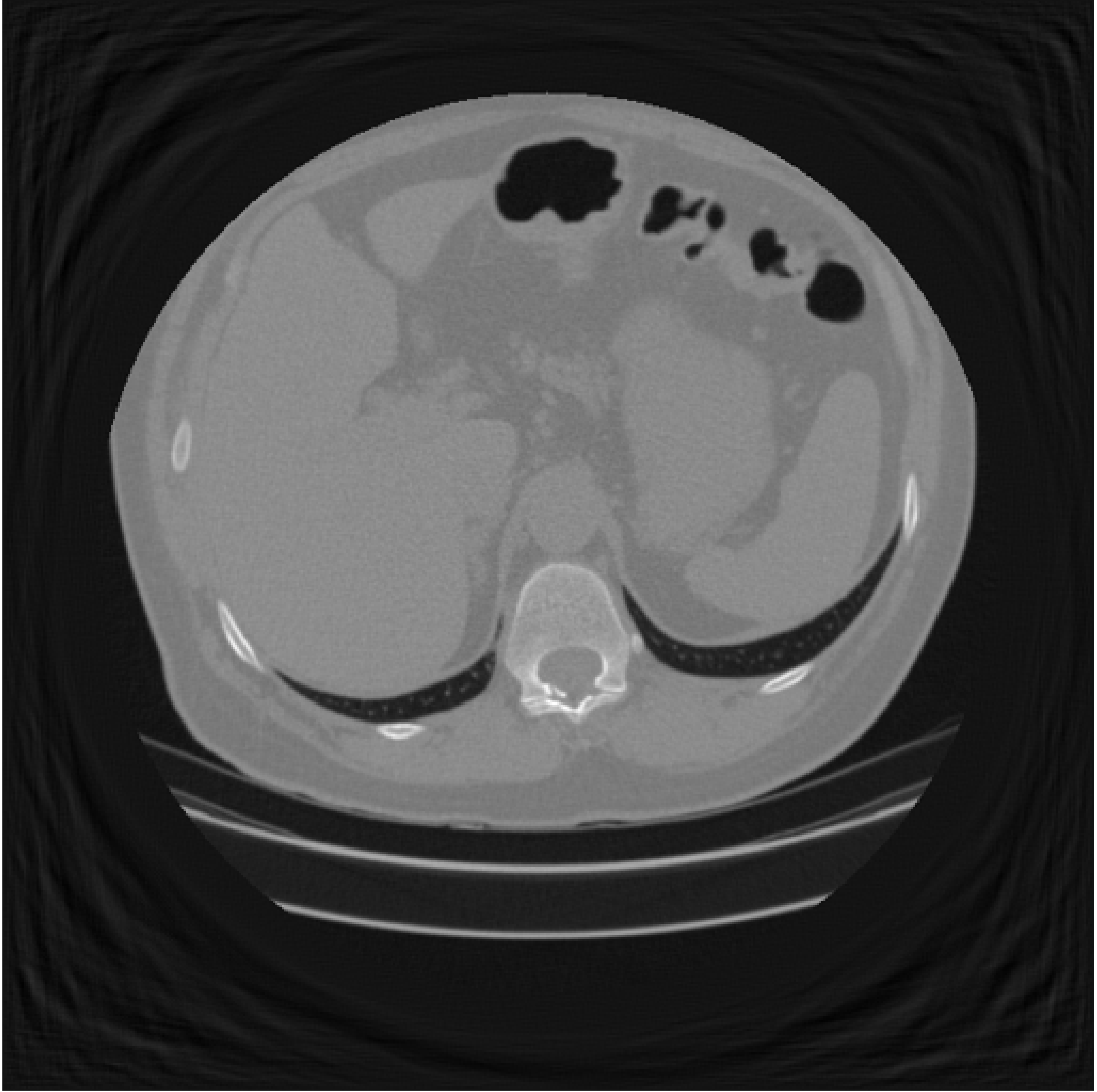} \\
  \end{tabular}
  \vspaceBefFigCaption
  \caption{Reference and the two reconstructed images
           of the slice studied 
           (slice with worst metrics in the entire collection).
           Left image, reference;
           center image, reconstruction with double-precision; 
           right image, reconstruction with single-precision.}
  \label{fig:rec}
\end{figure}

\begin{figure}[ht!]
  \centering
  \begin{tabular}{c}
  \includegraphics[width=0.60\linewidth]{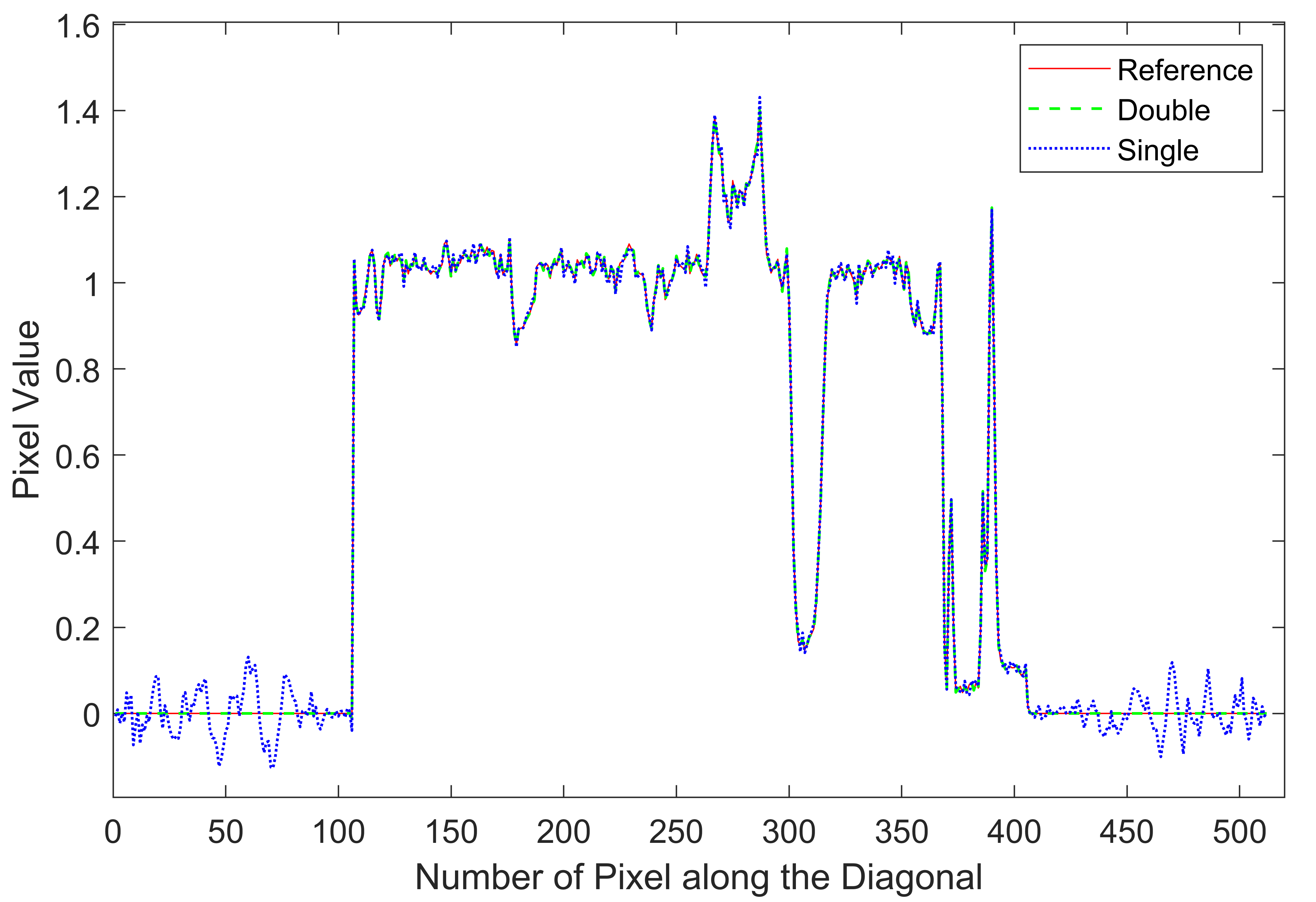} 
  \end{tabular}
  \vspaceBefFigCaption
  \caption{Profile of the pixel values 
           along the main diagonal of the slice studied
           (slice with worst metrics in the entire collection)
           for the three images: reference, double precision, and single precision.}
  \label{fig:profile}
\end{figure}

Figure~\ref{fig:rec} shows a reference image and 
their two corresponding reconstructed images
(one for double precision and one for single precision).
Concretely, the image processed is the one that 
obtains the worst quality metrics in the entire collection of 2048 images.
For double precision, the values of SNR, PSNR, and SSIM 
are $195.16$, $206.1$, and $1$, respectively.
For single precision, they are $28.11$, $38.27$, and $0.99994$, respectively.
A close observation of the reconstructed images 
may notice 
(easier on the screen than on paper)
that in the single-precision image 
the largest differences or errors appear to be outside the body.
To visualize this much better, 
Figure \ref{fig:profile} shows the profile of pixel values 
along the diagonal of the image of Figure \ref{fig:rec} (top-left to bottom-right). 
It can be seen that the highest errors occur in the outer areas 
in which the reference values are equal to 0. 

\begin{figure}[ht!]
  \centering
  \begin{tabular}{cc}
    \includegraphics[width=0.30\linewidth]{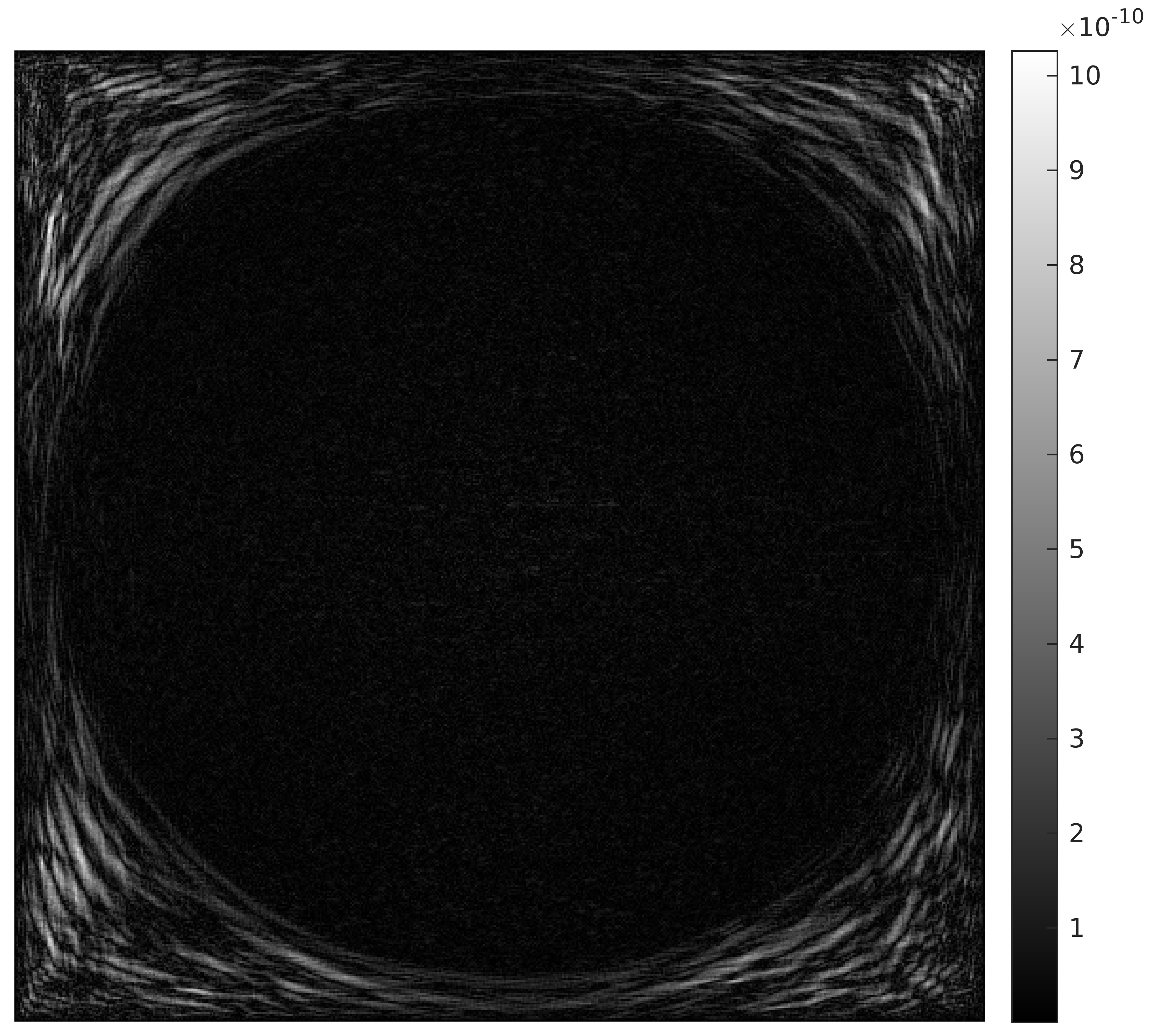} &
    \includegraphics[width=0.30\linewidth]{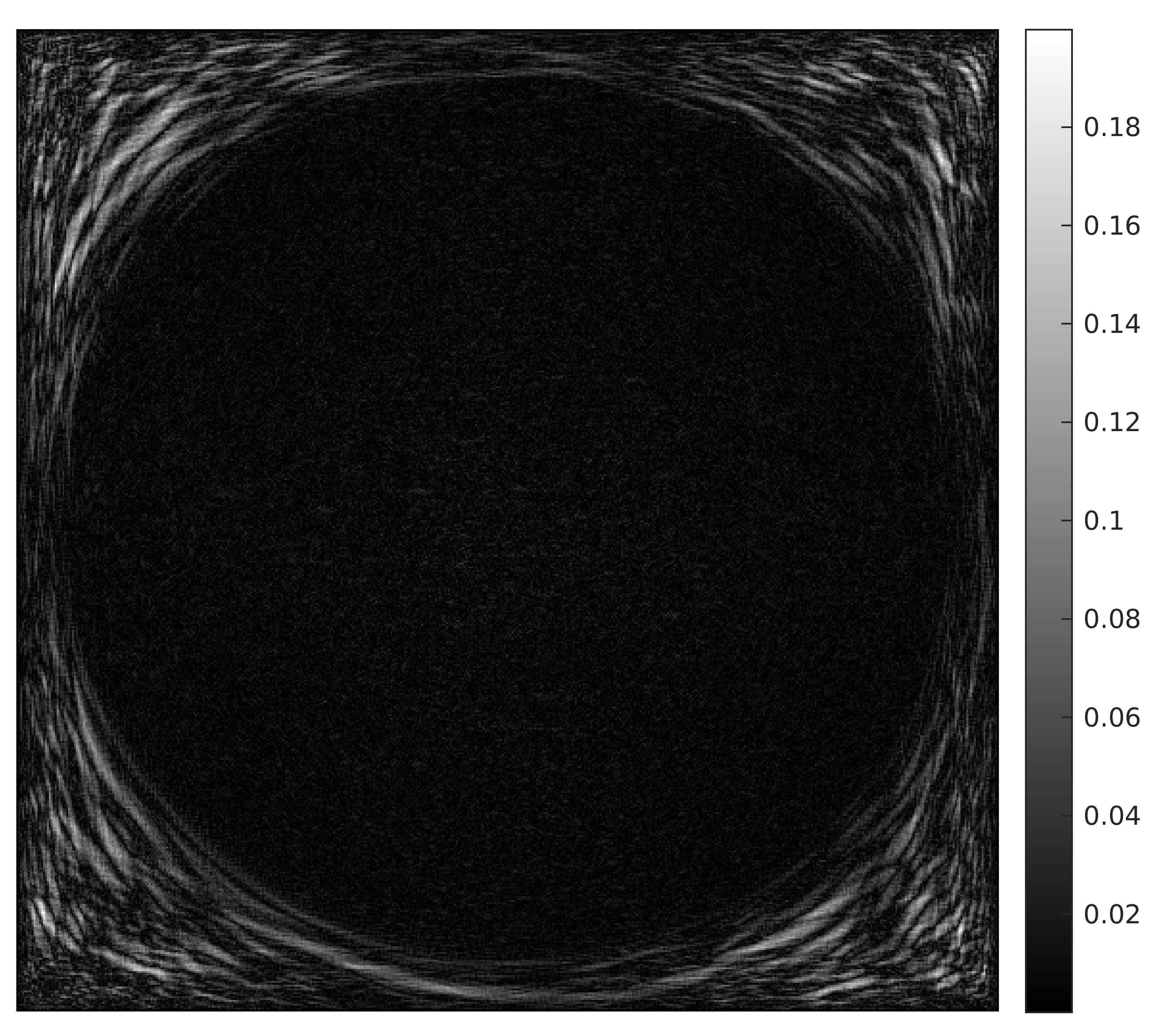} \\
  \end{tabular}
  \vspaceBefFigCaption
  \caption{Errors in the reconstructed images
           of the slice studied
           (slice with worst metrics in the entire collection).
           Left, double-precision; right, single-precision.}
  \label{fig:diff}
\end{figure}

Figure~\ref{fig:diff} shows the images 
obtained by subtracting the reconstructed image and the reference image
for both double-precision and single-precision arithmetic.
Clearly, the largest errors 
are obtained in the outer parts of the image.
In other words, 
the error mainly affects the areas with just air, outside the patient's body.
This is the reason why the PSNR is higher than the SNR, 
since the highest intensity values do not show as much noise 
as the lowest intensity values.
%
%
Observing the range of the images, 
it can be seen that the maximum error in the pixel values 
obtained with double precision is about $10^{-9}$, 
while for single precision it is about $10^{-1}$, which is a large difference. 
Nevertheless, the internal structures are not visually altered by noise, 
at least for the human eye. 
For this reason, the SSIM is very high and very close to 1.

\begin{figure}[ht!]
  \centering
  \begin{tabular}{c}
  \includegraphics[width=0.30\linewidth]{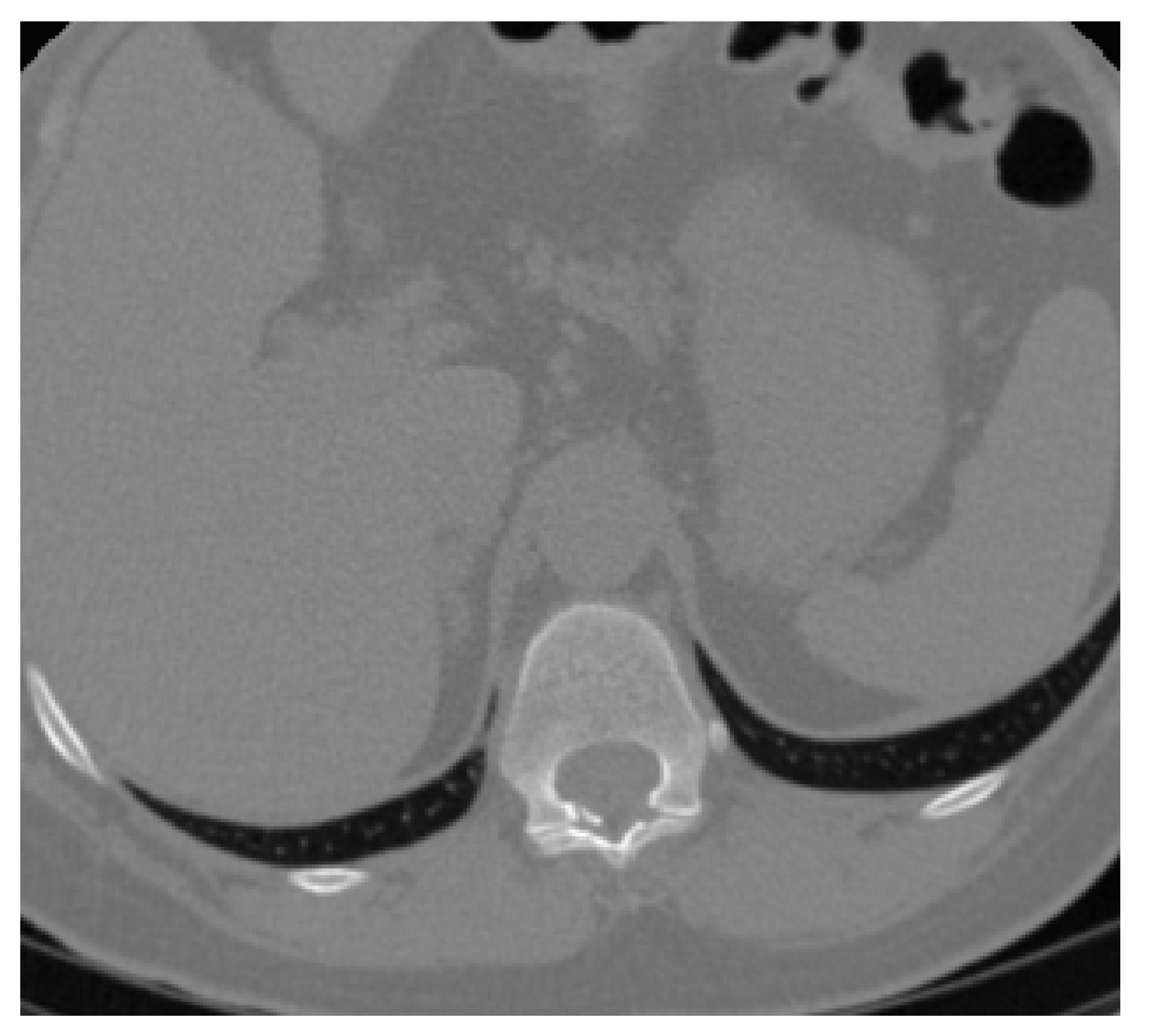}
  \end{tabular}
  \vspaceBefFigCaption
  \caption{Cropped image}
  \label{fig:cropped}
\end{figure}

\begin{table}[ht!]
  \centering
  \begin{tabular}{|c|c|c|c|} \hline
  \textbf{Precision} & \textbf{SNR} & \textbf{PSNR} & \textbf{SSIM} \\ \hline
  Double             &  209.84      & 219.27        & 1         \\ \hline
  Single             &   42.74      &  52.18        & 0.9999993 \\ \hline
  \end{tabular}
  \vspaceBefTableCaption
  \caption{Mean image quality of the 2048 slices when cropping the edges.}
  \label{tab:quality_cropped}
\end{table}

Figure~\ref{fig:cropped} 
shows the resulting image when cropping the outer parts
of the previous image:
130 pixels at the bottom edge and 
100 pixels at the other edges (left, top, and right).
Table \ref{tab:quality_cropped} shows the values of the metrics obtained in this case.  
The average SNR has improved by approximately 11 points for single precision
and 12 for double precision. 
The average PSNR has improved by approximately 8 points for single precision
and 9 for double precision. 
The average SSIM for single precision falls short of 1, 
since the noise has not been completely removed, 
but the noise is so small that the human eye cannot perceive the difference. 

\begin{figure}[ht!]
  \centering
  \begin{tabular}{c}
  \includegraphics[width=0.7\linewidth]{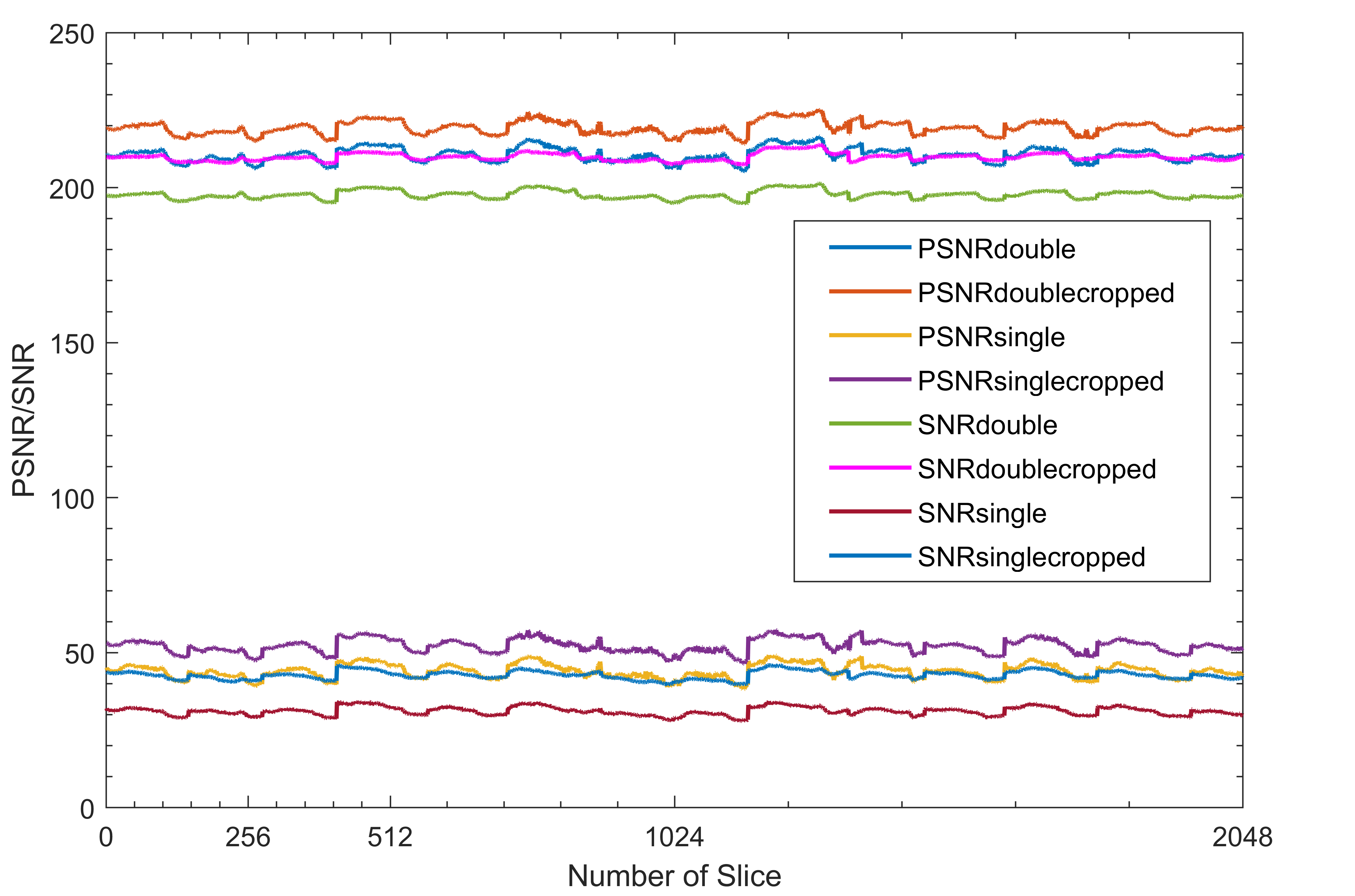} 
  \end{tabular}
  \vspaceBefFigCaption
  \caption{PSNR and SNR of the 2048 slices for every variant.}
  \label{fig:metrics}
\end{figure}

Figure \ref{fig:metrics} 
shows the SNR and PSNR metrics for all the 2048 slices of the four reconstructions:
both non-cropped and cropped for both double precision and single precision.

To conclude on the results of this dataset, 
we can state that the visual differences 
between the reference images and the reconstructed images are negligible
despite the fact that the reconstructed images 
obtained with single-precision arithmetic contain a higher level of noise.
Although the reconstructions can no longer be considered perfect, 
as was the case with double precision due to its very small error, 
it can still be said that they are high-quality reconstructions 
on par with many other reconstruction methods.

\subsection{Dataset DICOM-CT-PD}

Another interesting area of application of our algebraic reconstruction method 
is the reconstruction of low-dose CT images. 
For this purpose, the DICOM-CT-PD dataset, developed by the Mayo Clinic, 
is of great relevance. 
This dataset contains around 300 CT studies of real patients, 
classified by the following body areas: 
abdomen, thorax, and brain~\citep{DICOM-CT-PD}.
Remarkably, each study provides both the reconstructed images and 
their corresponding projections 
for three CT scanners from different manufacturers. 
In addition, although in a simulated form, 
the low-dose projections and their reconstructions are also provided. 
To simulate the low-dose acquisition, 
routine dose (full-dose) acquisitions 
using standard clinical protocols 
for the anatomical region of interest are used. 
Then, Poisson noise is added to each projection dataset 
to create a second projection dataset with a lower simulated dose level. 
This error simulation technique accurately models 
the impact of tube current modulation,  bowtie filtering, and electronic noise.
The head and abdomen cases are provided at 25\% of the routine dose, and 
the thorax cases are provided at 10\% of the routine dose. 
For this part of our study, 
the thorax case C016 of the dataset has been selected, 
since the thorax cases are the scans with the largest dose reduction. 

\subsubsection{Simulated Data}

In this case, the generation of the input data has been performed 
with the ASTRA Toolbox~\citep{astra1, astra2},
a widely used library for advanced tomographic reconstructions 
and simulations of medical imaging, 
especially in Computed Tomography. 
It was designed with a focus on efficiency and flexibility, 
offering tools for both sinogram simulation and image reconstruction. 
It also provides several 2D and 3D geometries and 
reconstruction methods. 

By using this toolbox, 
new sinograms have been calculated using both the full-dose and low-dose images. 
Then, the coefficient matrix $A$ for the reconstruction problem has been generated. 
The original sinograms provided by the dataset have not been used 
due to two reasons:
First, the projection data was acquired with helical-beam CT scanners, 
and our QR method works with axial fan-beam projections. 
Second, the dataset contains data from different manufacturers,
and thus different numbers of detectors and configurations. 
By re-projecting the images using ASTRA, 
we select a standard configuration 
so that our implementation can process any study from the dataset 
and compare the results. 

Table~\ref{tab:scanner2} details the parameters of the configuration used,
very similar to the previous case. 
The numbers of projections are different for reasons explained below.
The resolution of the images will also be increased 
in order to test the stability of the reconstruction method. 

\begin{table}[ht!]
\centering
  \begin{tabular}{|c|c|c|} \hline
    \textbf{Parameter}                & \multicolumn{2}{|c|}{\textbf{Value}} \\  \hline
    Source trajectory                 &  \multicolumn{2}{|c|}{$360^\textrm{o}$ circular scan } \\  \hline
    Source-to-isocenter distance      & \multicolumn{2}{|c|}{65 cm} \\  \hline
    Source-to-detector distance       & \multicolumn{2}{|c|}{150 cm }\\  \hline
    X-ray source fan angle            & \multicolumn{2}{|c|}{$33^\textrm{o}$} \\  \hline
    Number of detectors               & \multicolumn{2}{|c|}{1025} \\  \hline
    Pixels of the reconstructed image & $512 \times 512$ & $768 \times 768$ \\  \hline
    Number of projections             & 264, 270 & 590, 610, 620 \\ \hline
  \end{tabular}
  \vspaceBefTableCaption
  \caption{\small{Simulated fan-beam scanner parameters.}}
  \label{tab:scanner2}
\end{table}

In the previous dataset
we have studied the general behavior of the reconstruction method 
for routine-dose (full-dose) scans
using the standard image resolution of $512\times512$ pixels. 
However, in some cases it can be very interesting
to reduce the radiation dose,
to increase the image resolution, or both at the same time.
With the current dataset 
we will analyze the full-dose and low-dose results 
of the C016 patient for the standard $512 \times 512$ image resolution 
as well as a higher resolution of $768\times768$ pixels. 
Since the QR reconstruction method can suffer from rounding errors, 
the resolution of a very high-dimensionality problem 
could result in higher residuals and lower-quality solutions. 
We aim to assess the performance of the method in this situation 
in terms of image quality.

\subsubsection{Quality of images with $512 \times 512$ pixels}

Now, the results for the reconstructed images 
with $512 \times 512$ pixels are shown and analyzed.
Unlike with the previous dataset, 
both full dose and low dose will be analyzed.

It is important to note that in our analysis for the previous dataset,
we checked that single precision with 260 projections 
was enough to provide a high quality in the reconstructions.
With this dataset, 
we need to employ a few projections more just because of the different
data generation method, and not due to our reconstruction method.
Although the algorithm for generating the input data 
(matrix $A$ and the sinograms) for the reconstructions is the same 
(a forward-projection algorithm based on Joseph's method), 
the implementation with ASTRA is different and 
a few more projections are necessary.
To further extend our analysis, 
we have employed two different number of projections (264 and 270) 
instead of only one (260).

\begin{figure}[ht!]
  \centering
  \begin{subfigure}[h]{0.48\linewidth}
    \centering
    \includegraphics[trim={1.0cm 0.8cm 1.0cm 1.0cm},clip,width=0.97\linewidth]{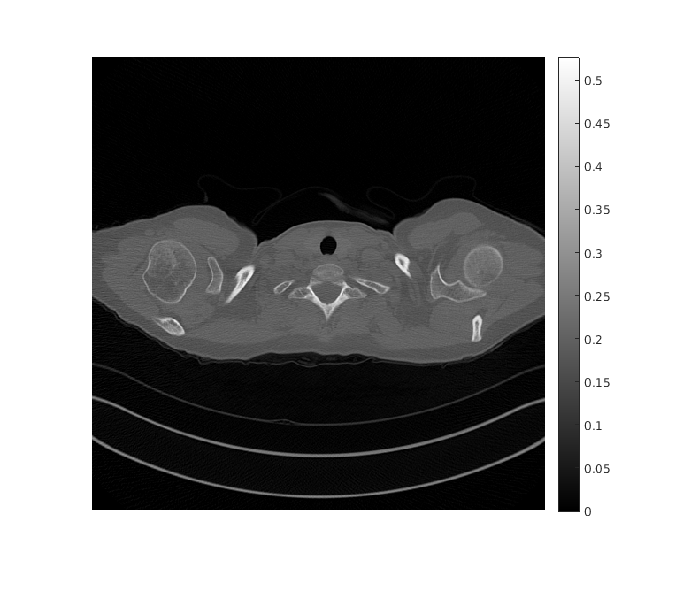}
    \vspace*{-1.0cm}
    \subcaption{Reference.}
  \end{subfigure}
  \begin{subfigure}[h]{0.48\linewidth}
    \centering
    \includegraphics[trim={1.0cm 0.8cm 1.0cm 1.0cm},clip,width=0.97\linewidth]{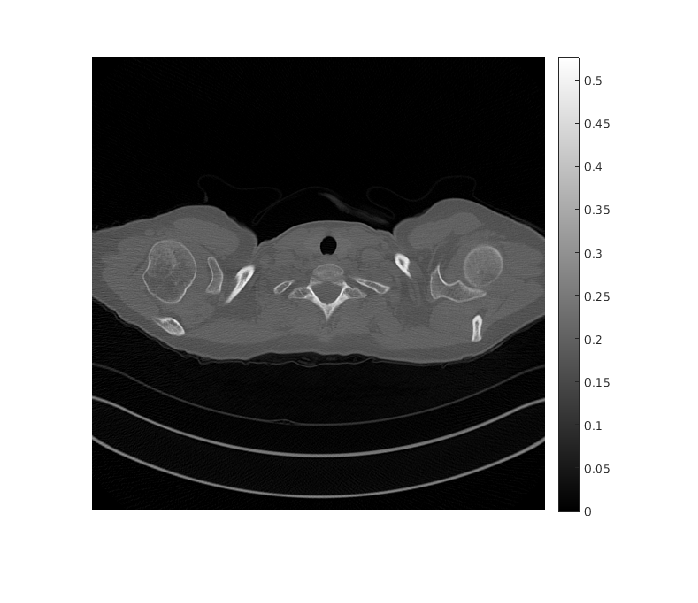}
    \vspace*{-1.0cm}
    \subcaption{Double precision and 264 projections}
  \end{subfigure} 
  \begin{subfigure}[h]{0.48\linewidth}
    \centering
    \includegraphics[trim={1.0cm 0.8cm 1.0cm 1.0cm},clip,width=0.97\linewidth]{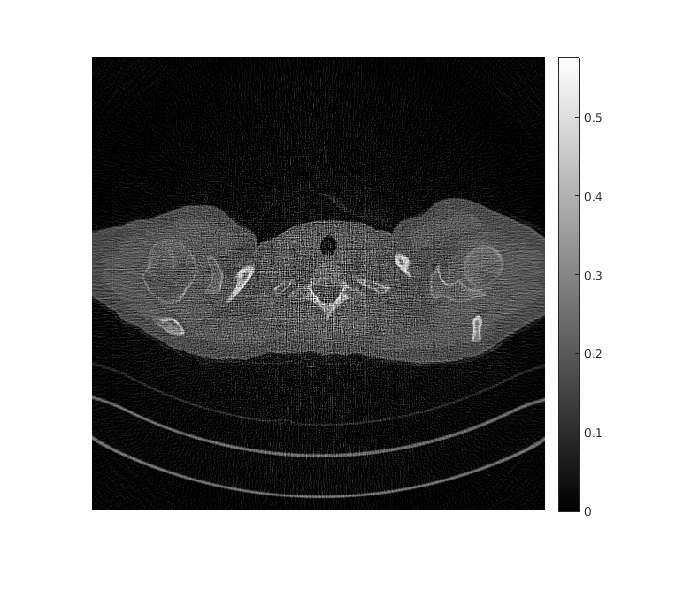}
    \vspace*{-1.0cm}
    \subcaption{Single precision and 264 projections}
  \end{subfigure}
  \begin{subfigure}[h]{0.48\linewidth}
    \centering
    \includegraphics[trim={1.0cm 0.8cm 1.0cm 1.0cm},clip,width=0.97\linewidth]{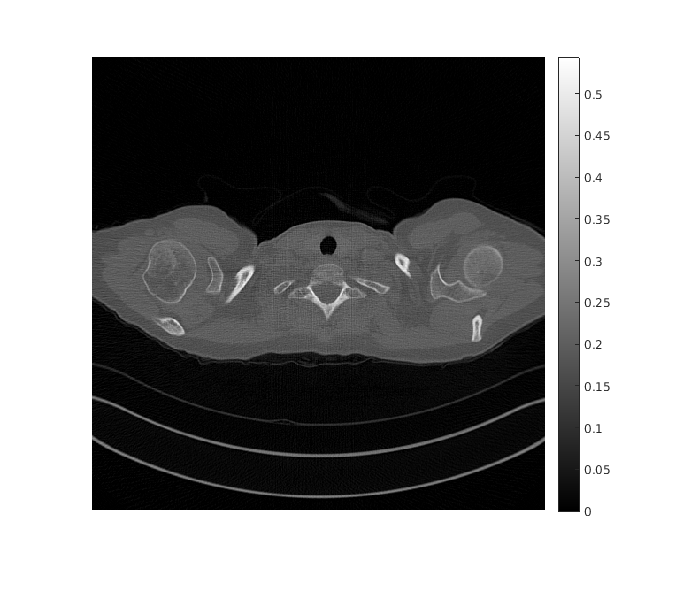}
    \vspace*{-1.0cm}
    \subcaption{Single precision and 270 projections}
  \end{subfigure}
  \vspace{-0.2cm}
  \caption{Reference and the three reconstructed images 
           with $512 \times 512$ pixels and full dose.}
  \label{fig:rec2}
\end{figure}

Figure~\ref{fig:rec2} shows 
the reference and the three reconstructed images of slice 10 of patient C016
with $512 \times 512$ pixels and full dose.
The top left image is the reference one.
The top right image is the image reconstructed with double-precision arithmetic.
The two bottom images are those reconstructed with single precision 
and two different number of projections: 264 (botom left) and 270 (bottom right).
As can be seen,
a high quality solution is obtained
when working with double precision and 264 projections.
In contrast, the single-precision image with 264 projections 
contains significant noise as well as reconstruction artifacts. 
Using only a few more projections (270),
the quality obtained becomes very similar to the double-precision solution.
Table \ref{tab:residual2} shows the residuals of the resolution 
for the reconstruction of this image for both precisions. 
The single-precision residual is in the same range as in the previous study.

\begin{table}[ht!]
  \centering
  \begin{tabular}{|c|c|c|} \hline
    \textbf{Precision} &  \textbf{\# Projections} & \textbf{Residual} \\ \hline
    Double  & 264  & $7.93 \cdot 10^{-15}$ \\ \hline
    Single  & 270  & $7.44 \cdot 10^{-6}$   \\ \hline
  \end{tabular}
  \vspaceBefTableCaption
  \caption{Residuals $\| AX-B \|_F / \| B \|_F$ of $AX=B$
           for $512 \times 512$ pixels and full dose.}
  \label{tab:residual2}
\end{table}

\begin{figure}[t!]
  \centering
  \begin{subfigure}[h]{0.48\linewidth}
    \centering
    \includegraphics[trim={1.0cm 0.8cm 1.0cm 1.0cm},clip,width=0.97\linewidth]{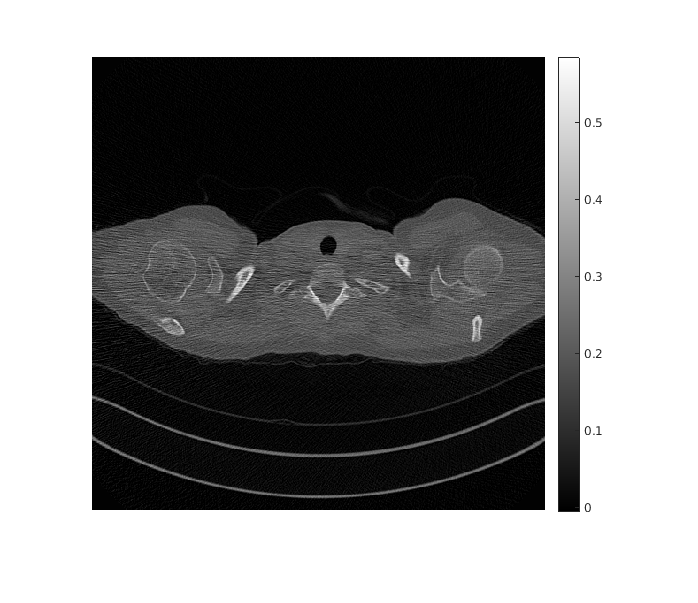}
    \vspace*{-1.0cm}
    \subcaption{Reference.}
  \end{subfigure}
  \begin{subfigure}[h]{0.48\linewidth}
    \centering
    \includegraphics[trim={1.0cm 0.8cm 1.0cm 1.0cm},clip,width=0.97\linewidth]{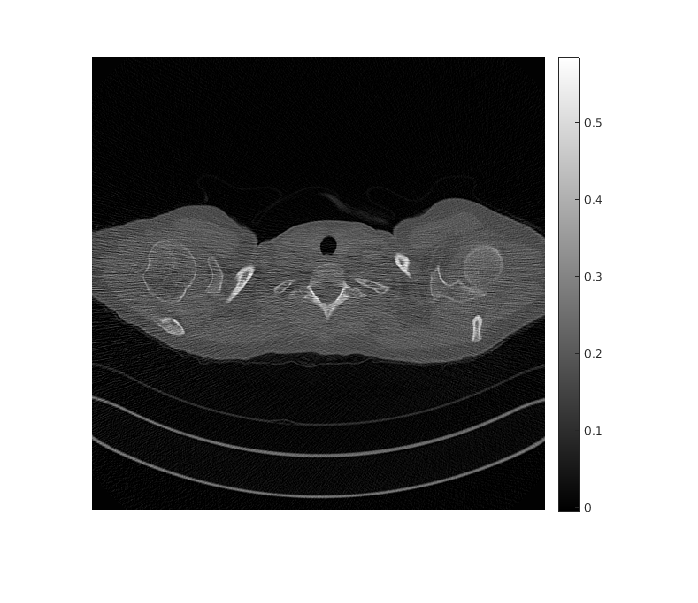}
    \vspace*{-1.0cm}
    \subcaption{Double precision and 264 projections}
  \end{subfigure}
  \begin{subfigure}[h]{0.48\linewidth}
    \centering
    \includegraphics[trim={1.0cm 0.8cm 1.0cm 1.0cm},clip,width=0.97\linewidth]{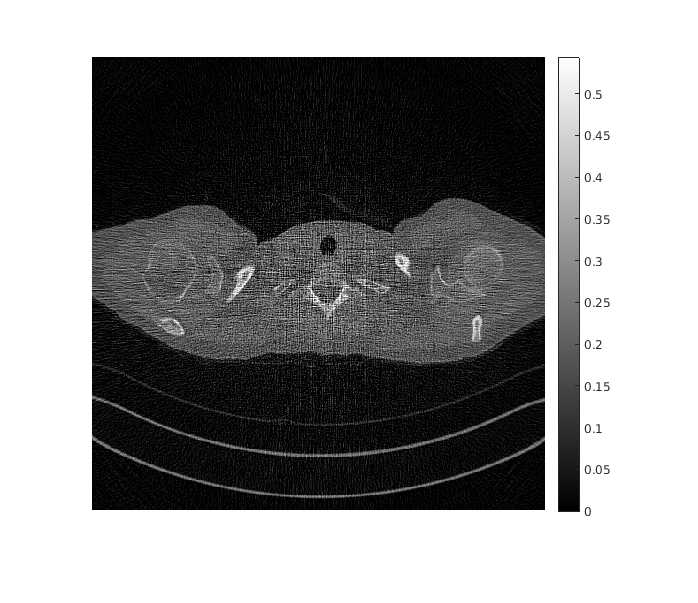}
    \vspace*{-1.0cm}
    \subcaption{Single precision and 264 projections}
  \end{subfigure}
  \begin{subfigure}[h]{0.48\linewidth}
    \centering
    \includegraphics[trim={1.0cm 0.8cm 1.0cm 1.0cm},clip,width=0.97\linewidth]{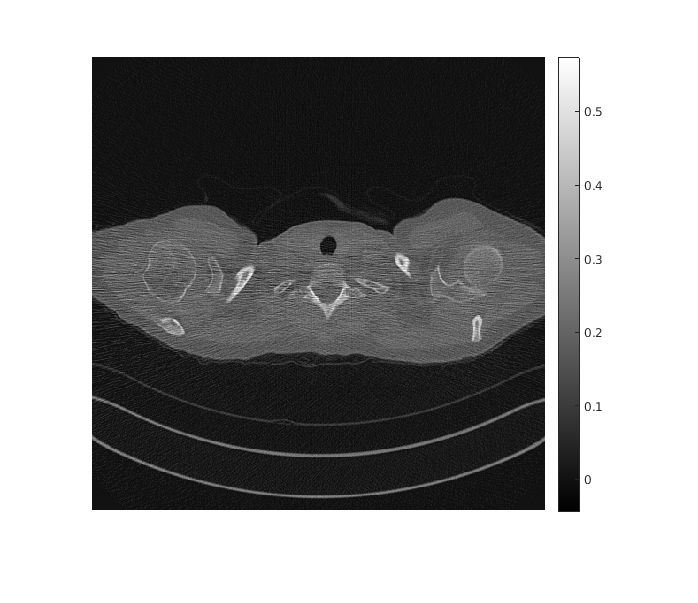}
    \vspace*{-1.0cm}
    \subcaption{Single precision and 270 projections}
  \end{subfigure}
  \vspace{-0.2cm}
  \caption{Reference and the three reconstructed images 
           with $512 \times 512$ pixels and low dose.}
  \label{fig:reclow}
\end{figure}

Figure~\ref{fig:reclow} includes the analogous images of the same slice
for the low-dose case.
As can be seen, 
the reference image obtained with low dose is not as clean as the full-dose one, 
since it even contains streak artifacts. 
Nevertheless, that is the image we aim to obtain with the reconstruction method, 
since it is the one that has been projected. 
Once again, the reconstruction with double-precision data
is visually identical to the reference. 

Table~\ref{tab:metricas} details the values of the PSNR and SSIM metrics 
of all cases with $512 \times 512$ pixels:
full dose and low dose of the three reconstructions.
As can be seen, 
the dose in the reconstructions is not so important 
since the quality of both reconstructions is almost identical. 
Therefore, the most impacting factors in the metrics 
are the arithmetic precision and the number of projections. 
Comparing the values of the PSNR metric, they have a wide range:
Double-precision reconstructions obtain 206 and 207,
single-precision reconstructions with 264 projections obtain 24 and 25, whereas
single-precision reconstructions with 270 projections obtain 38 and 39.
Comparing the values of the SSIM metric,
the values of all the reconstructions stay very close to $1$.
In this case, the noise in the image is not as perceptible by the human eye. 

\begin{figure}[ht!]
  \centering
  \begin{subfigure}[h]{0.33\linewidth}
    \includegraphics[trim={3cm 1.0cm 3.8cm 1.0cm},clip,width=0.99\linewidth]{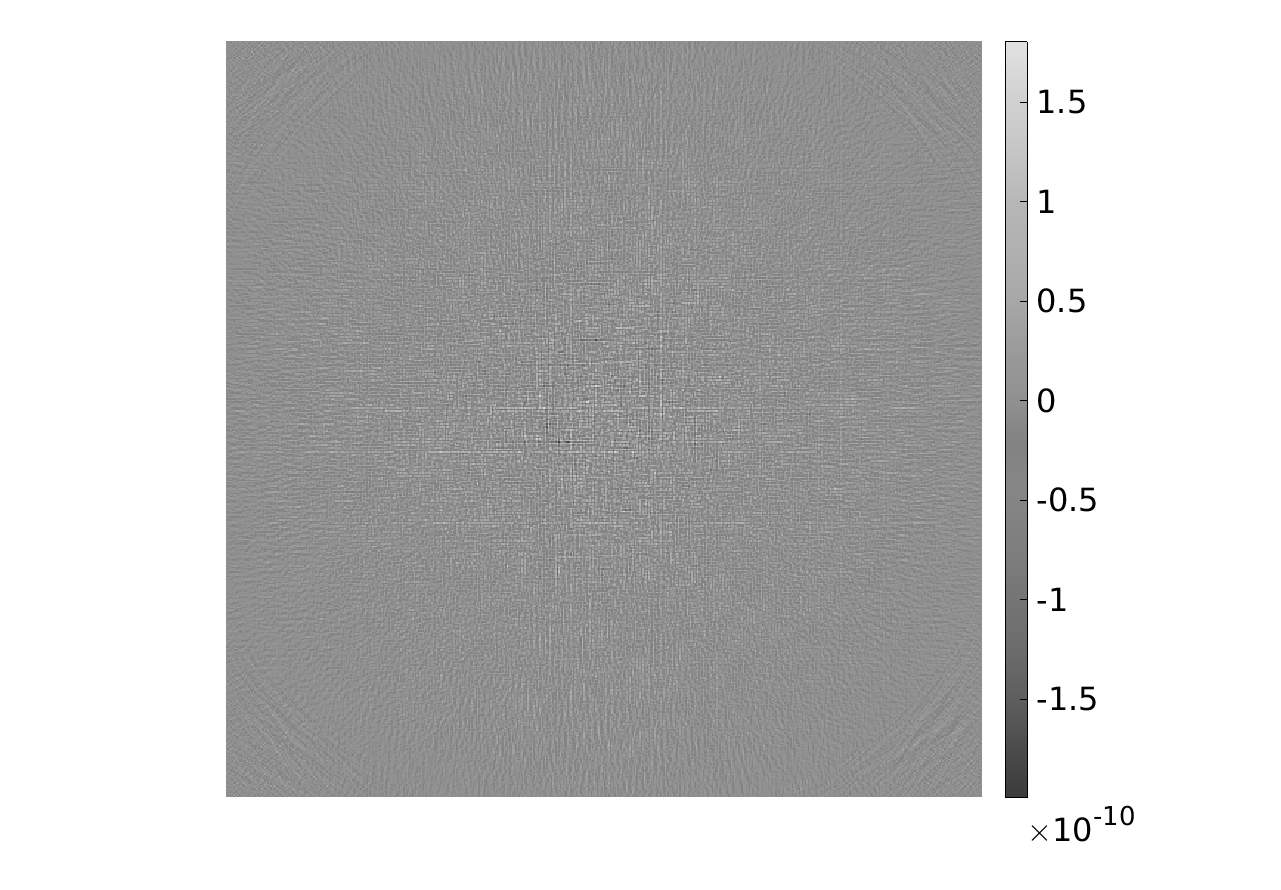}
  \end{subfigure}
  \begin{subfigure}[h]{0.32\linewidth}
    \includegraphics[trim={3cm 1.0cm 4cm 1.0cm},clip,width=0.99\linewidth]{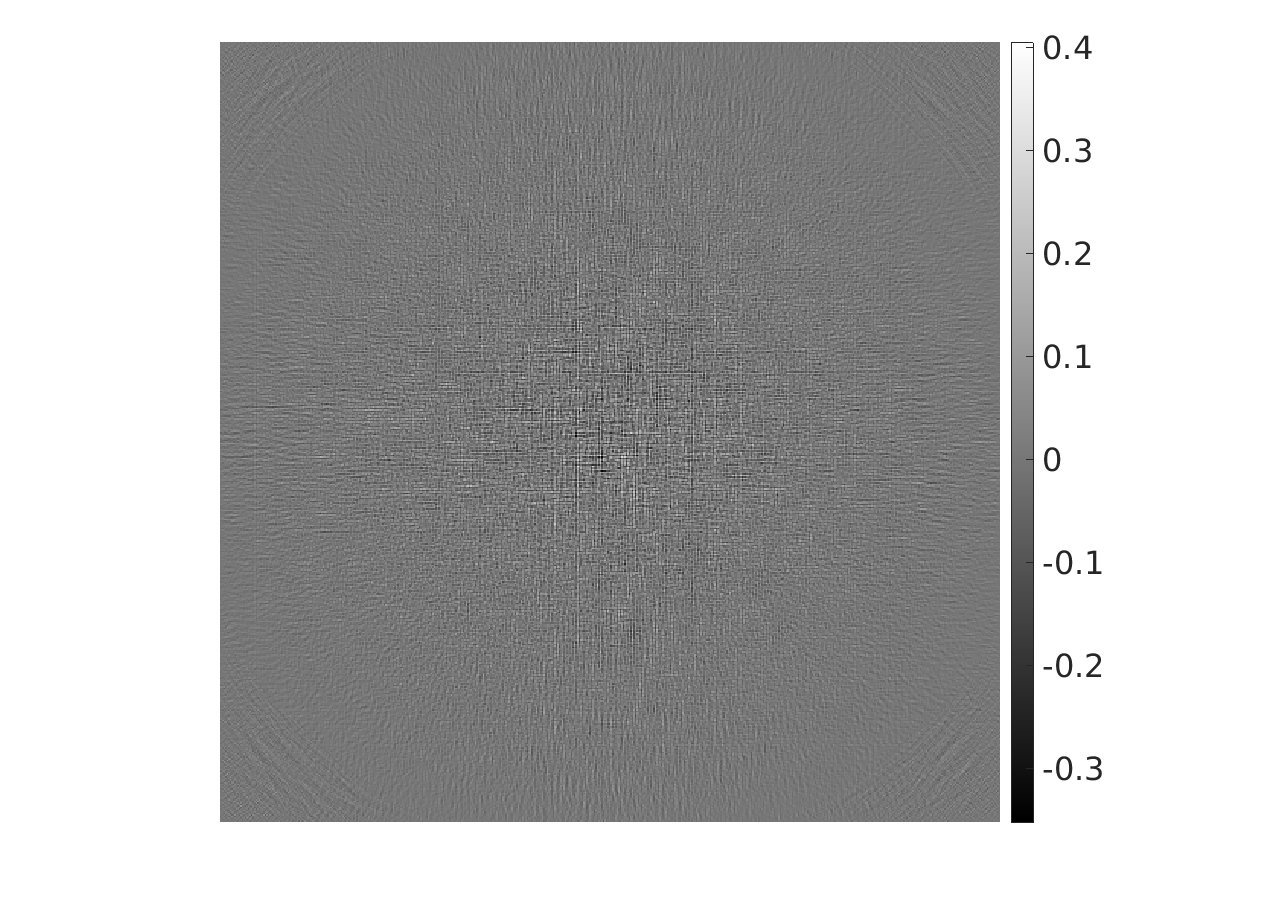}
    \end{subfigure}
   \begin{subfigure}[h]{0.32\linewidth}
    \includegraphics[trim={3cm 1.0cm 4cm 1.0cm},clip,width=0.99\linewidth]{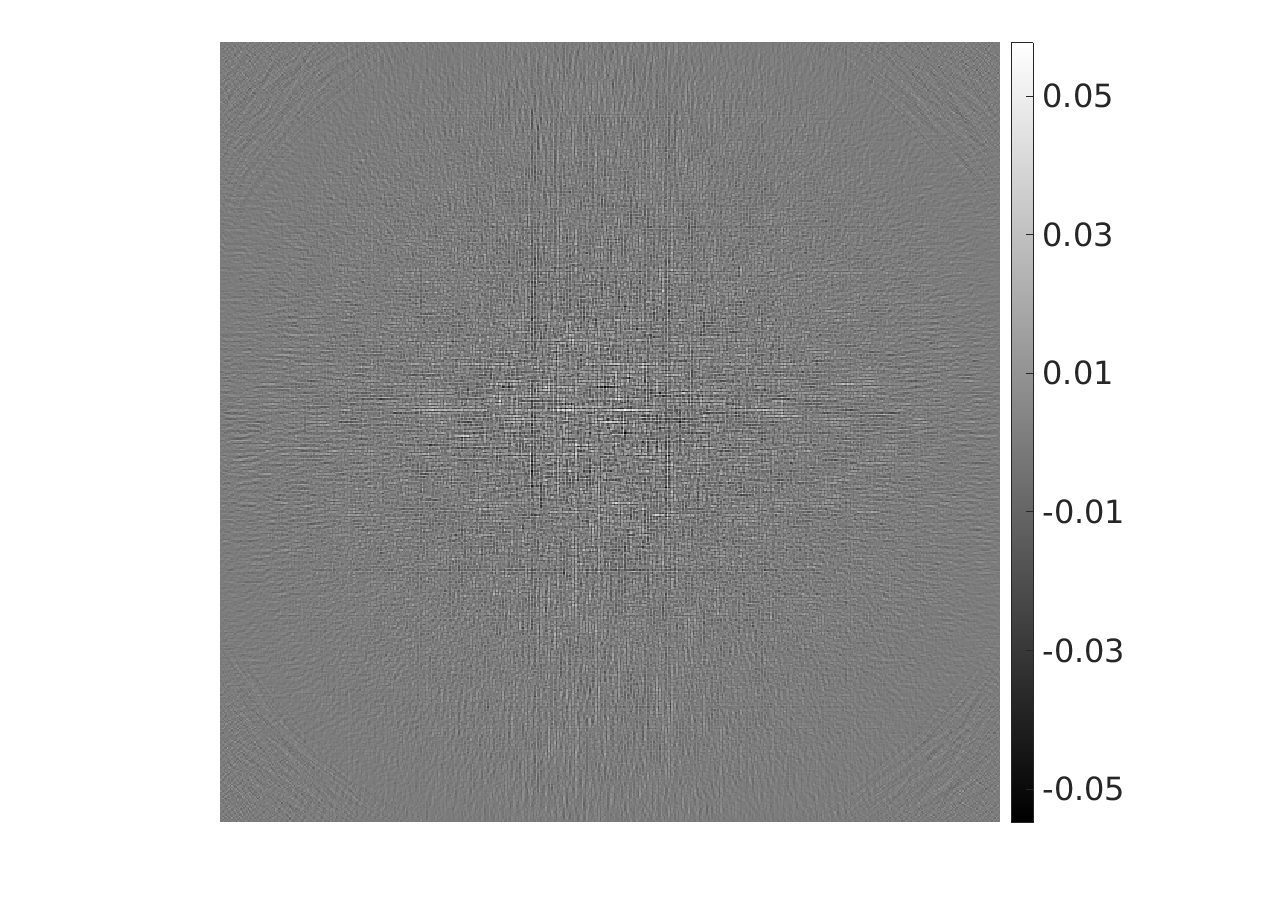}
  \end{subfigure}
  \vspaceBefFigCaption
  \caption{Error images of the reconstructions for $512 \times 512$ pixels and low dose.
  Left, error of double precision and 264 projections;
  center, error of single precision and 264 projections;
  right, error of single precision and 270 projections.
  }
  \label{fig:errorlowdose512}
\end{figure}

Figure~\ref{fig:errorlowdose512} shows the difference or error 
of the three reconstructed images with low dose 
compared to the reference one.
The maximum error is about $10^{-10}$ for double-precision images, 
it is about 0.4 for single precision and 264 projections, 
and it drops to about 0.06 when the number of projections is increased to 270.
It can be concluded that the single-precision reconstruction method 
still gets valid results even though the number of projections needs to be slightly higher 
when using the ASTRA toolbox to calculate the forward projection.

\begin{table}[ht!]
  \centering
  \vspace{0.2cm}
  \begin{tabular}{|c|c|c|c|c|c|} \hline
    \textbf{Reference} & 
    \textbf{Reconstr.~projec.} & 
    \textbf{Precision} & 
    \textbf{\# Projec.} & 
    \textbf{PSNR} & 
    \textbf{SSIM} \\ \hline
    \multirow{3}{*}{\begin{tabular}{c}Full Dose\\ FBP\end{tabular}}
       & Full dose  & Double  & 264  & 206  & 1         \\ \cline{2-6} 
       & Full dose  & Single  & 264  & 24   & 0.99997   \\ \cline{2-6} 
       & Full dose  & Single  & 270  & 38   & 0.9999992 \\ \hline
    \multirow{3}{*}{\begin{tabular}{c}Low Dose\\ FBP\end{tabular}}
       & Low dose   & Double  & 264  & 207  & 1         \\ \cline{2-6} 
       & Low dose   & Single  & 264  & 25   & 0.99997   \\ \cline{2-6} 
       & Low dose   & Single  & 270  & 39   & 0.9999992 \\ \hline
  \end{tabular}
  \vspaceBefTableCaption
  \caption{Image quality of both full-dose and low-dose results 
           for both single and double precision.}
  \label{tab:metricas}
\end{table}

\subsubsection{Quality of images with $768 \times 768$ pixels}

Now, the results for the reconstructed images 
with $768 \times 768$ pixels are shown and analyzed.
Both full dose and low dose are assessed.

In clinical practice, it can be very interesting
to increase the image resolution in some cases to obtain a finer detail.
Note that the $768 \times 768$ resolution contains 2.25 times as many pixels 
as the standard $512 \times 512$ resolution.
Increasing the resolution of the reconstructed images 
implies that the size of the problem also increases. 
This will require to perform more computations 
which could result in higher residuals and lower-quality solutions. 
The goal is to assess the performance of the method in this situation 
in terms of image quality with both full dose and low dose.

To reconstruct images of size $768 \times 768$, 
the minimum number of projections needed to get a full-rank matrix $A$ 
is 576, but 590 have been taken to ensure good-quality solutions. 
Thus, the matrix dimension of $A$ is $604,750 \times 589,824$. 
Recall that when working with $512 \times 512$ pixels and single precision
a few more projections were used because of the data generation method
for this dataset.
In this case we also decided to increase the number of projections 
to 610 and 620 for single precision,
with matrix dimensions $625,250 \times 589,824$ and 
$635,500 \times 589,824$ respectively.

\begin{figure}[b!]
  \centering
  \begin{subfigure}[h]{0.48\linewidth}
    \centering
    \includegraphics[trim={1.0cm 0.8cm 1.0cm 1.0cm},clip,width=0.97\linewidth]{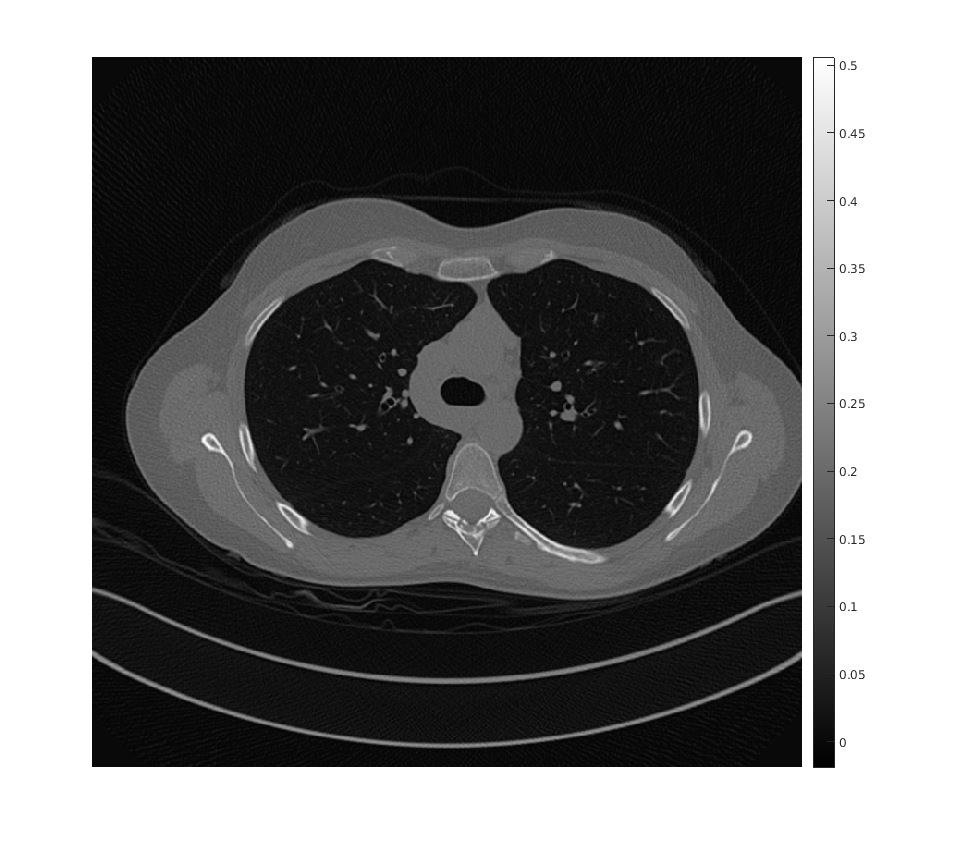}
    \vspace*{-0.7cm}
    \subcaption{Reference.}
  \end{subfigure}
  \begin{subfigure}[h]{0.48\linewidth}
    \centering
    \includegraphics[trim={1.0cm 0.8cm 1.0cm 1.0cm},clip,width=0.97\linewidth]{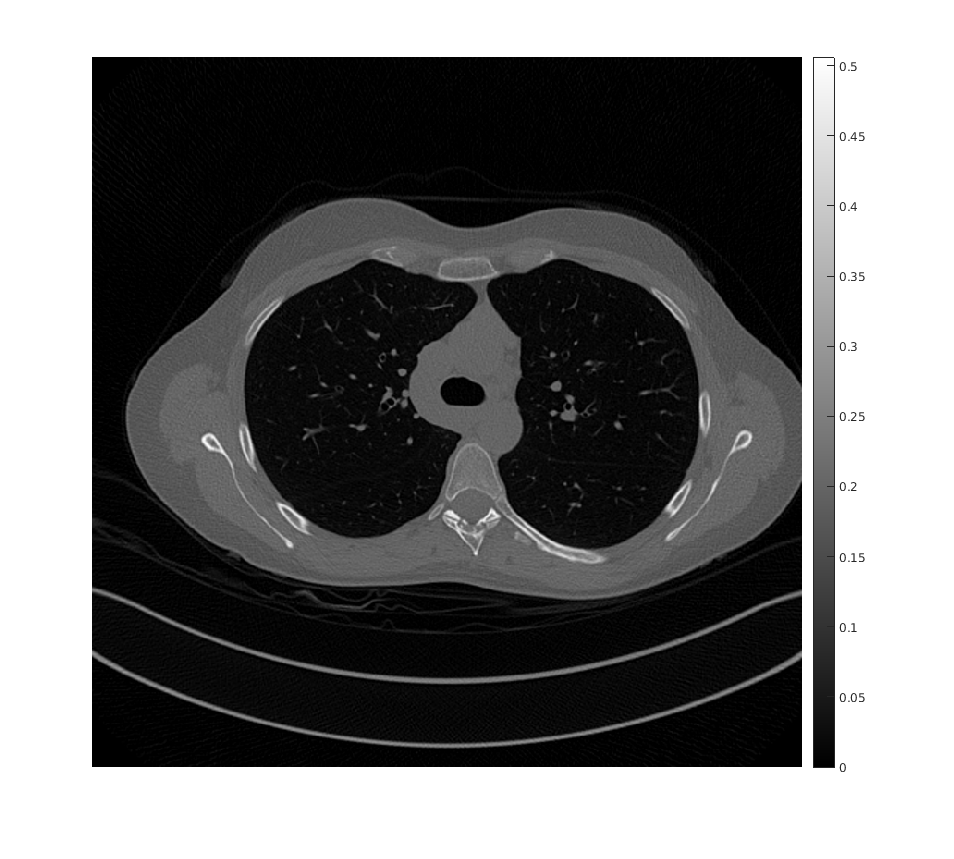}
    \vspace*{-0.7cm}
    \subcaption{Double precision and 590 projections.}
  \end{subfigure}
  \begin{subfigure}[h]{0.48\linewidth}
    \centering
    \includegraphics[trim={1.0cm 0.8cm 1.0cm 1.0cm},clip,width=0.97\linewidth]{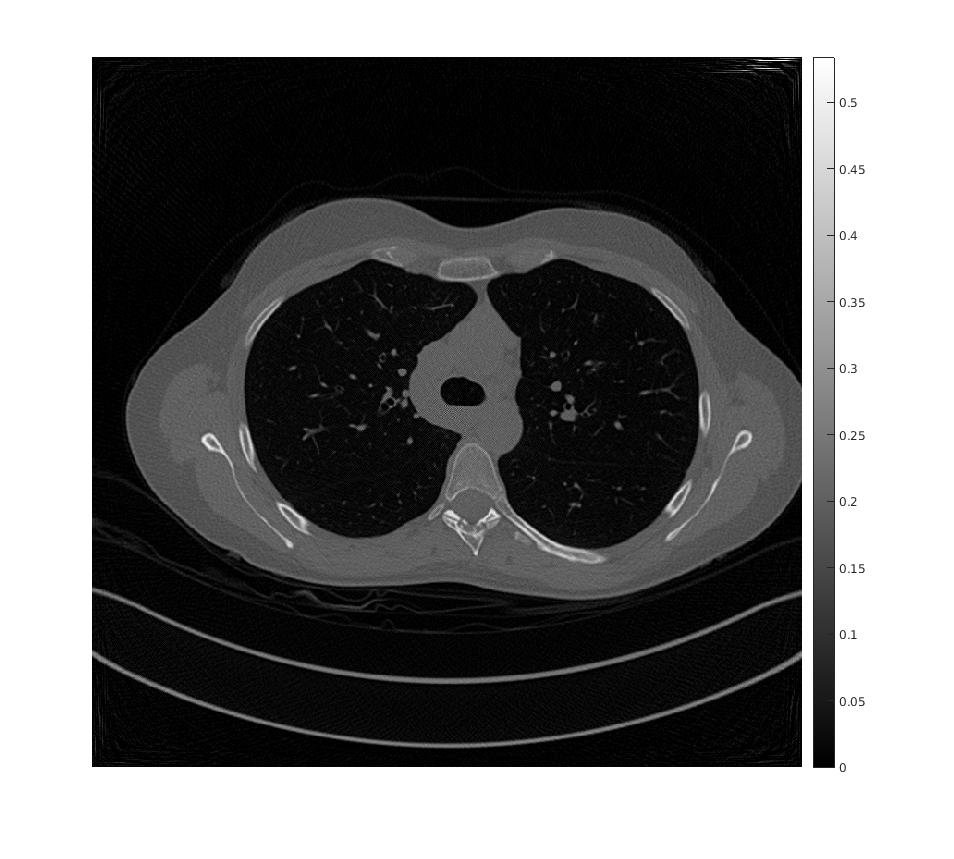}
    \vspace*{-0.7cm}
    \subcaption{Single precision and 610 projections.}
  \end{subfigure}
  \begin{subfigure}[h]{0.48\linewidth}
    \centering
    \includegraphics[trim={1.0cm 0.8cm 1.0cm 1.0cm},clip,width=0.97\linewidth]{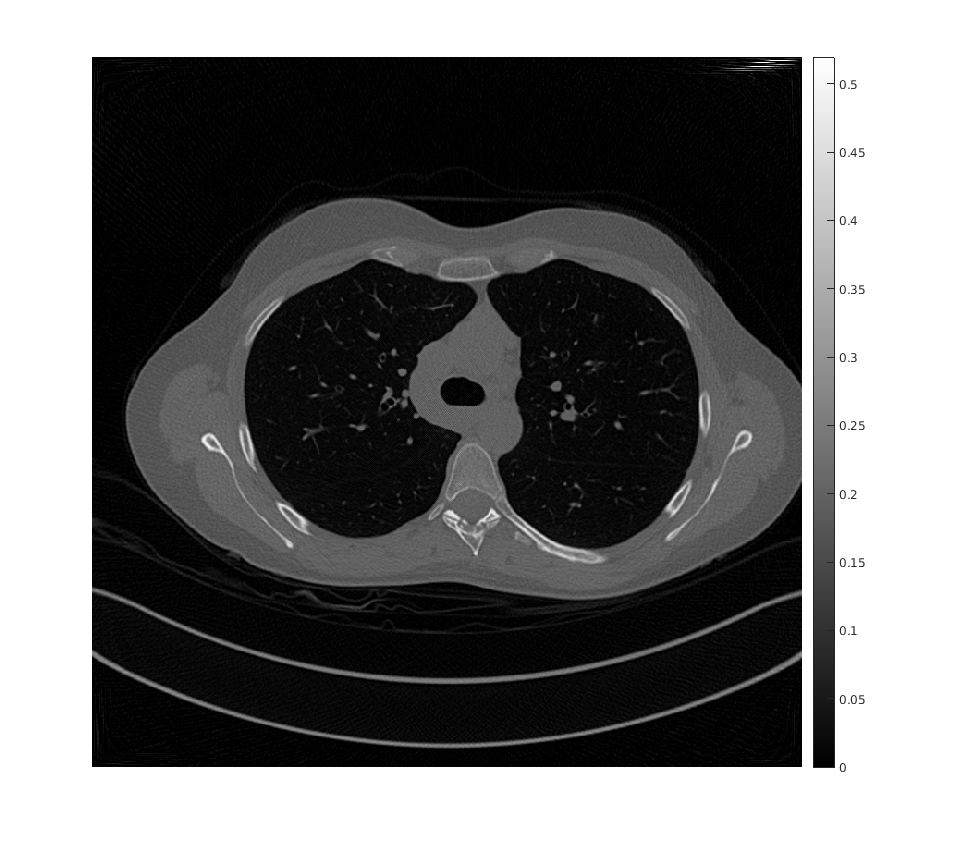}
    \vspace*{-0.7cm}
    \subcaption{Single precision and 620 projections.}
  \end{subfigure}
  \vspace{-0.2cm}
  \caption{Reference and the three reconstructed images 
           with $768 \times 768$ pixels and full dose.}
  \label{fig:rec3}
\end{figure}

Figure~\ref{fig:rec3} shows 
the reference and the three reconstructed images of slice 100 of patient C016
with the increased resolution $768 \times 768$ and full dose.
The top left image is the reference one.
The top right image is the image reconstructed with double-precision arithmetic.
The two bottom images are those reconstructed with single precision 
and two different number of projections: 610 (bottom left) and 620 (bottom right).
The single-precision reconstruction with 610 projections
shows some artifacts in the corners.
With 620 projections, the artifacts are mostly gone, 
except for the top-right corner. 
The artifacts appear when using single-precision arithmetic and 
the number of projections is similar to the number needed for double precision. 
Nevertheless, the quality of the result with 620 projections is high enough and 
would be a good compromise between image quality and dose reduction.

Table~\ref{tab:residual3} compares the residuals. 
Obviously, they are slightly higher (worse) than for the $512 \times 512$ resolution. 
The residual for the single-precision solution is 
larger than $10^{-6}$,  which is the usual target, but not too much.
However, the image has high quality.

\begin{table}[t!]
  \centering
  \begin{tabular}{|c|c|c|} \hline
    \textbf{Precision} &  \textbf{\# Projections} & \textbf{Residual}   \\ \hline
    Double  & 590  & $1.01 \cdot 10^{-14}$ \\ \hline
    Single  & 620  & $1.70 \cdot 10^{-5}$ \\ \hline
  \end{tabular}
  \vspaceBefTableCaption
  \caption{Residuals $\| AX-B \|_F / \| B \|_F$ of $AX=B$
           for $768\times 768$ pixels and full dose.}
  \label{tab:residual3}
\end{table}

\FloatBarrier

\begin{figure}[t!]
  \centering
  \begin{subfigure}[h]{0.48\linewidth}
    \centering
    \includegraphics[trim={1.0cm 0.8cm 1.0cm 1.0cm},clip,width=0.97\linewidth]{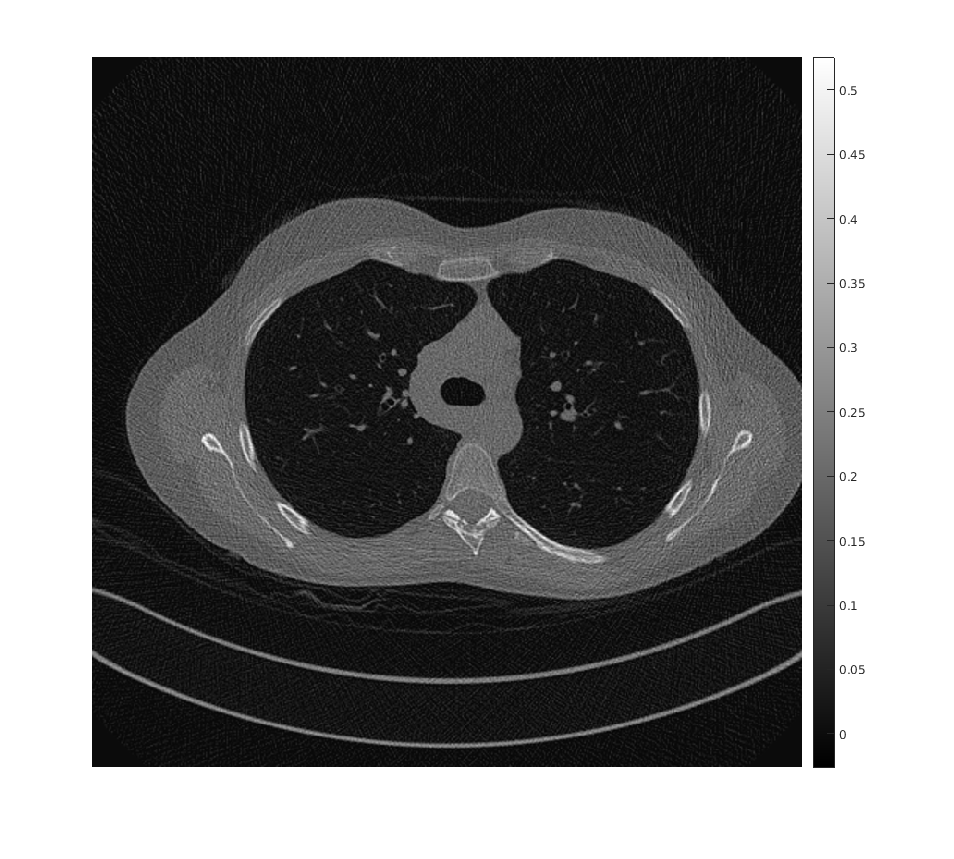}
    \vspace*{-0.7cm}
    \subcaption{Reference.}
  \end{subfigure}
  \begin{subfigure}[h]{0.48\linewidth}
    \centering
    \includegraphics[trim={1.0cm 0.8cm 1.0cm 1.0cm},clip,width=0.97\linewidth]{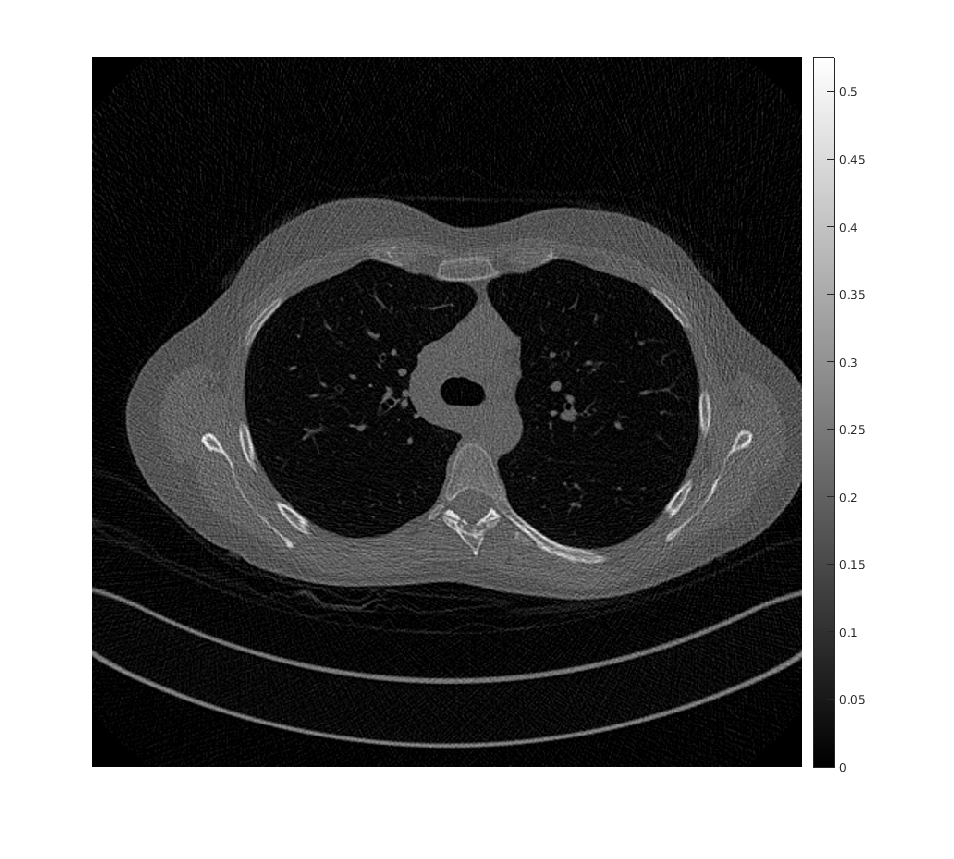}
    \vspace*{-0.7cm}
    \subcaption{Double precision and 590 projections.}
  \end{subfigure}
  \begin{subfigure}[h]{0.48\linewidth}
    \centering
    \includegraphics[trim={1.0cm 0.8cm 1.0cm 1.0cm},clip,width=0.97\linewidth]{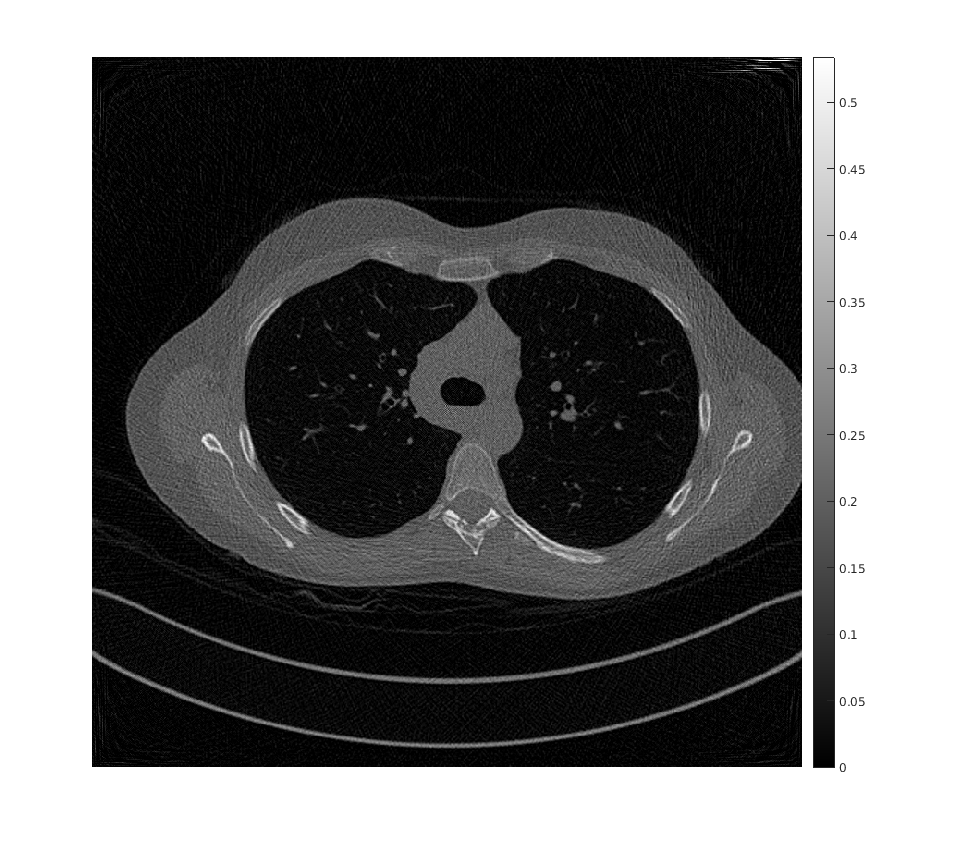}
    \vspace*{-0.7cm}
    \subcaption{Single precision and 610 projections.}
  \end{subfigure}
  \begin{subfigure}[h]{0.48\linewidth}
    \centering
    \includegraphics[trim={1.0cm 0.8cm 1.0cm 1.0cm},clip,width=0.97\linewidth]{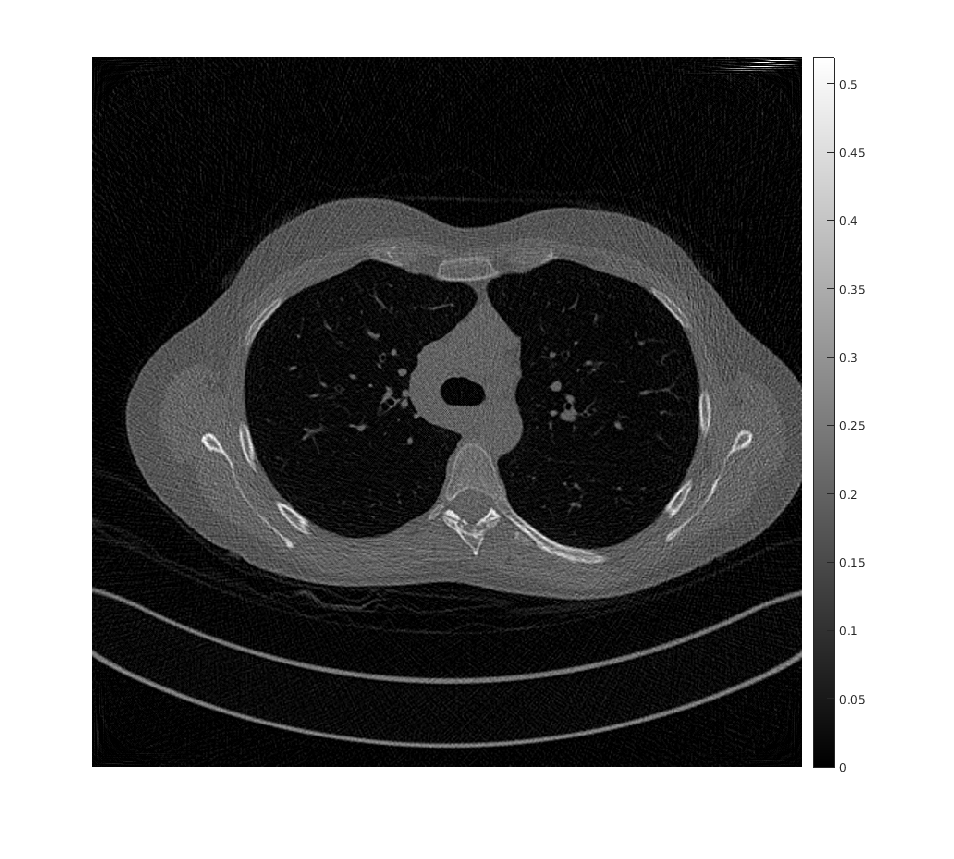}
    \vspace*{-0.7cm}
    \subcaption{Single precision and 620 projections.}
  \end{subfigure}
  \vspace{-0.2cm}
  \caption{Reference and the three reconstructed images 
           with $768 \times 768$ pixels and low dose.}
  \label{fig:reclow2}
\end{figure}

\begin{figure}[ht!]
  \centering
  \begin{subfigure}[h]{0.33\linewidth}
    \centering
    \includegraphics[trim={1.0cm 1.2cm 0.3cm 0.27cm},clip,width=0.97\linewidth]{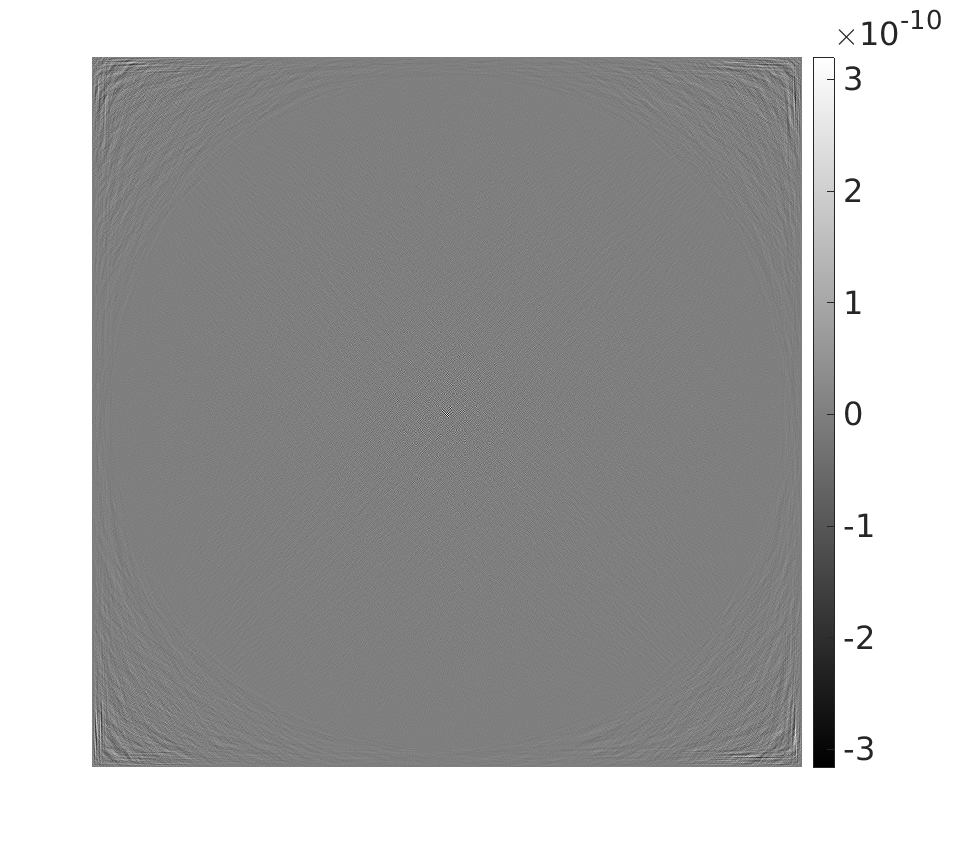}
  \end{subfigure}
  \begin{subfigure}[h]{0.32\linewidth}
    \centering
    \includegraphics[trim={1.0cm 1.2cm 0.7cm 0.3cm},clip,width=0.97\linewidth]{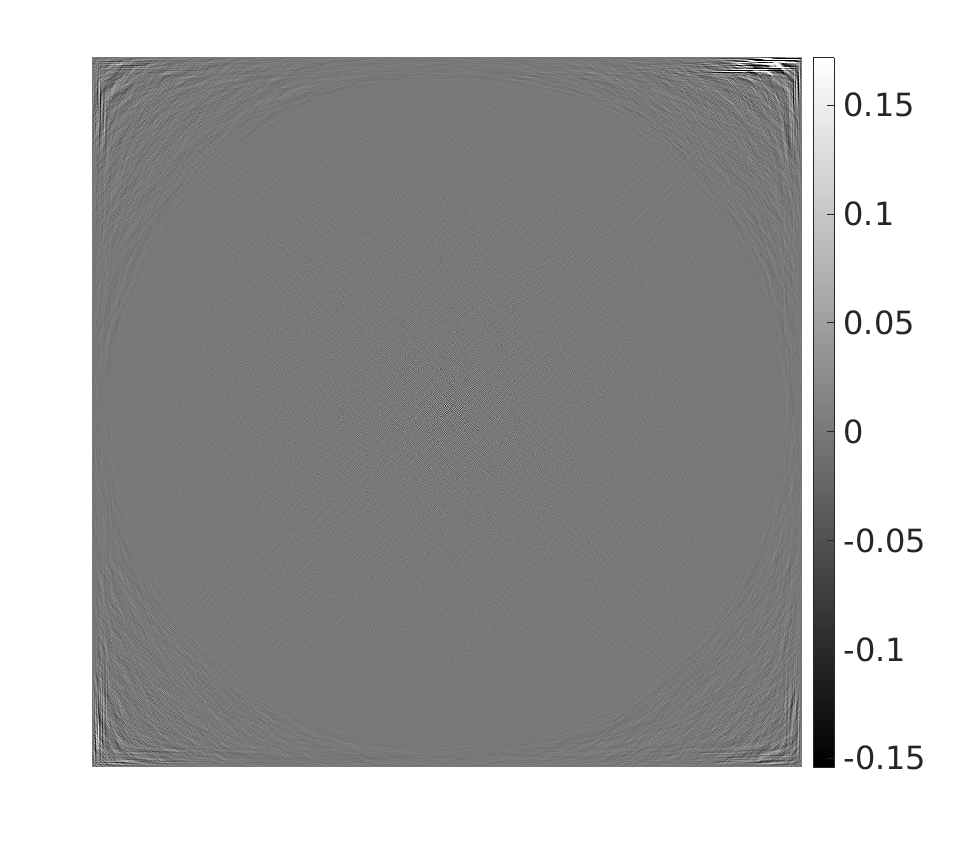}
  \end{subfigure}
  \begin{subfigure}[h]{0.32\linewidth}
    \centering
    \includegraphics[trim={1.0cm 1.2cm 0.7cm 0.3cm},clip,width=0.97\linewidth]{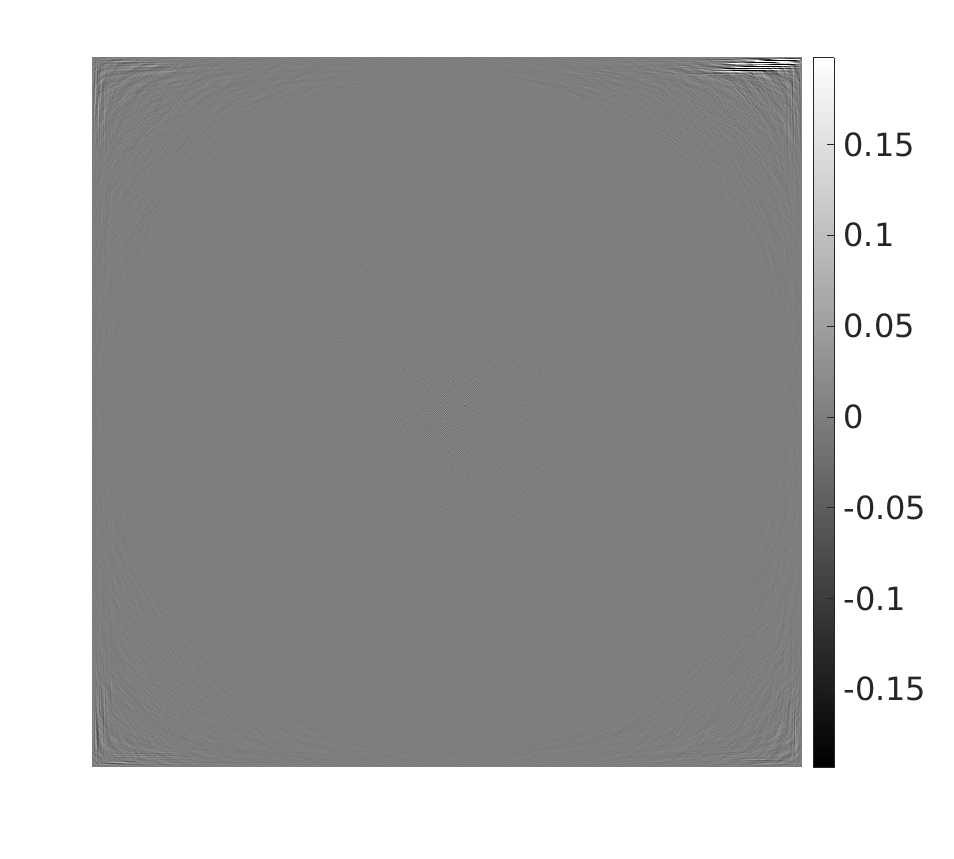}
  \end{subfigure}
  \vspaceBefFigCaption
  \caption{Error images of the reconstructions for $768 \times 768$ pixels and low dose.
  Left, error of double precision and 590 projections;
  center, error of single precision and 610 projections;
  right, error of single precision and 620 projections.
  }
  \label{fig:error2}
\end{figure}

Figure~\ref{fig:reclow2} shows the corresponding results for low dose.
There is no visual difference between the double and single-precision results
except for some corners of the images. 
Figure~\ref{fig:error2} shows the difference between the reference and 
the reconstructed images with low dose.
The single-precision image with 610 projections 
contains noise in the corners as well as some smaller noise in the center.
The single-precision image with 620 projections 
removes all the noise at the center
and reduces the noise at the corners.

Table~\ref{tab:metricas2} contains 
the quality metrics for all high-resolution reconstructions.
The PSNR and SSIM with 610 and 620 projections are not very different,
unlike the results in Table~\ref{tab:metricas}. 
That is because 610 is already a higher number of projections 
compared to the double-precision case. 
All single-precision images have a higher level of pixel noise, 
as shown by the PSNR results. 
Recall that for the standard $512 \times 512$ image resolution, 
the maximum PSNR obtained was 39. 
For the increased resolution $768 \times 768$, 
the maximum PSNR obtained is 34. 
However, the SSIM values are very close to 1, 
which proves that the images have good quality and 
the internal structures are well-preserved. 
All in all, these results verify the validity of the QR reconstruction method 
for high-resolution full-dose and low-dose CT images.

\begin{table}[ht!]
  \centering
  \vspace{0.2cm}
  \begin{tabular}{|c|c|c|c|c|c|} \hline
    \textbf{Reference} & 
    \textbf{Reconstr.~projec.} & 
    \textbf{Precision} & 
    \textbf{\# Projec.} & 
    \textbf{PSNR} & 
    \textbf{SSIM} \\ \hline
    \multirow{3}{*}{\begin{tabular}{c}Full Dose\\ FBP\end{tabular}}
       & Full dose  & Double  & 590  & 209  & 1        \\ \cline{2-6} 
       & Full dose  & Single  & 610  & 32   & 0.999996 \\ \cline{2-6} 
       & Full dose  & Single  & 620  & 34   & 0.999998 \\ \hline
    \multirow{3}{*}{\begin{tabular}{c}Low Dose\\ FBP\end{tabular}}
       & Low dose   & Double  & 590  & 209  & 1        \\ \cline{2-6} 
       & Low dose   & Single  & 610  & 32   & 0.999996 \\ \cline{2-6} 
       & Low dose   & Single  & 620  & 34   & 0.999998 \\ \hline
  \end{tabular}
  \vspaceBefTableCaption
  \caption{Image quality of both full-dose and low-dose results 
           for both single and double precision 
           with a $768\times768$ resolution.}
  \label{tab:metricas2}
\end{table}

\section{Conclusions}

In this study, 
we have presented a new implementation of an algebraic method 
for the reconstruction of Sparse-view Computed Tomography (CT)
based on the dense QR factorization 
and single-precision floating-point arithmetic.

Moreover,
we have carried out a comparison of the QR method applied to CT medical image
reconstruction using both double-precision and single-precision arithmetic. 
Using single-precision arithmetic,
the time required to reconstruct a complete study with 2048 images 
on a last-generation GPU (NVIDIA A100) has been divided by about two.
This time improvement would mean obtaining a full body study in 76 seconds
with about $27$ reconstructed slices per second.
This speed is competitive with that obtained by manufacturers 
with their own optimized methods. 
This time and speed already places the QR method 
at the level of speed required for clinical practice.

On the other side,
the quality results obtained with single-precision arithmetic 
are high and competitive with other methods, 
but not as high as with double-precision arithmetic.
Working with double precision and $512 \times 512$ pixels,
the reconstructions obtained with the QR method 
are practically perfect due to their minimal error. 
Working with single precision, 
the noise level increases especially in the areas with air
(with pixel values close to 0).
However,
the structures corresponding to the patient's body are unaltered, 
despite the fact that in those areas there is also noise in the image. 
This noise is not perceptible to the human eye 
since it does not generate notable artifacts or distort structures, 
and therefore using single precision may be a viable option.

The results for high resolution contain more artifacts 
that can be mitigated increasing the number of projections acquired. 
But, since that would mean increasing the dose, 
a reasonable compromise should be reached.  


In future works, a more in-depth study 
about the quality of the data generated with ASTRA and 
the parameters of the simulated scanner will be performed. 
If matrix $A$ was improved, the number of projections needed could be decreased. 
In addition, several other CT datasets will be tested 
so that the images cover more body parts and 
have particular elements to look for, such as lesions or tumors. 
The results will be evaluated by medical professionals 
to assess the quality of the reconstructed images and 
to determine the validity of the proposed method.

\section*{Acknowledgments}

This research has been supported by ``Universitat Politècnica de València'', 
and  is part of the TED2021-131091B-I00 project, 
funded by MCIN/AEI/10.13039/501100011033 and 
by the ``European Union NextGenerationEU/PRTR''.


\bibliography{references}
\bibliographystyle{amsplain}

\end{document}